\documentclass[a4paper,5pt,twocolumn,preprint]{elsarticle}

   \usepackage{hyperref}
    \journal{Journal of \LaTeX\ Templates}
    
    \usepackage{hhline}  % http://ctan.org/pkg/hhline
    \usepackage{multirow}
    \usepackage{graphicx,booktabs}
    \usepackage{xfrac}
    \usepackage{titlesec}
    \usepackage{wrapfig,booktabs}
    \usepackage{mwe}
    \usepackage{adjustbox}
    \usepackage{caption}
    \usepackage{makecell}

    \usepackage[table]{xcolor}
    \usepackage{array}
    \usepackage{longtable}
    \usepackage{fullpage}
    \usepackage[tracking=true]{microtype}
    \usepackage{subfig}
    \usepackage[font=small,skip=0pt]{caption}

    \usepackage{amsmath,array}

    \usepackage[letterpaper, portrait, margin=0.65in]{geometry}
    \arrayrulecolor{red}
    \fboxsep=\tabcolsep\relax
    \fboxrule=\arrayrulewidth\relax

    \newcolumntype{A}[2]{%
    >{\minipage{\dimexpr#1\linewidth-2\tabcolsep-#2\arrayrulewidth\relax}\vspace\tabcolsep}%
    c<{\vspace\tabcolsep\endminipage}}

\begin{document}

\twocolumn[

{\begin{frontmatter}

    \title{On the Interfacial Phase Growth and Vacancy Evolution during Accelerated Electromigration in Cu/Sn/Cu Microjoints
    \tnotetext[mytitlenote]{Fully documented templates are available in the elsarticle package on \href{http://www.ctan.org/tex-archive/macros/latex/contrib/elsarticle}{CTAN}.}}

    %\title{Computational Investigation of Microstructure Formation and Evolution of Cu/Sn/Cu Joints under Operating Conditions in 3DIC Packaging Systems
    %\tnoteref{mytitlenote}
    %\tnotetext[mytitlenote]{Fully documented templates are available in the elsarticle package on \href{http://www.ctan.org/tex-archive/macros/latex/contrib/elsarticle}{CTAN}.}}

    \author{Vahid Attari$^{a,}$\fnref{myfootnote1}}
    \author{Supriyo Ghosh$^{a}$}
    \author{Thien Duong$^{a}$}
    \author{Raymundo Arroyave$^{a,b}$}
    \address{$^a$Materials Science and Engineering Department, Texas A\&M University, College Station,TX 77843}
    \address{$^b$Mechanical Engineering Department, Texas A\&M University, College Station,TX 77843}
    \cortext[mycorrespondingauthor]{Corresponding author email:attari.v@tamu.edu} 
     
    \begin{abstract} 
       In this work, we integrate different computational tools based on multi-phase-field simulations to account for the evolution of morphologies and crystallographic defects of Cu/Sn/Cu sandwich interconnect structures that are widely used in three dimensional integrated circuits (3DICs). Specifically, this work accounts for diffusion-driven formation and disappearance of multiple intermetallic phases during accelerated electromigration and takes into account the non-equilibrium formation of vacancies due to electromigration. The work compares nucleation, growth, and coalescence of intermetallic layers during transient liquid phase bonding and virtual joint structure evolution subjected to accelerated electromigration conditions at different temperatures. The changes in the rate of dissolution of Cu from intermetallics and the differences in the evolution of intermetallic layers depending on whether they act as cathodes or anodes are accounted for and are compared favorably with experiments. The model considers non-equilibrium evolution of vacancies that form due to differences in couplings between diffusing atoms and electron flows. This work is significant as the point defect evolution in 3DIC solder joints during electromigration has deep implications to the formation and coalescence of voids that ultimately compromise the structural and functional integrity of the joints.
    \end{abstract}


  \end{frontmatter}}
  
  ]
  
  %\linenumbers
  
  \section{Introduction}
    
    \fontdimen2\font=0.5ex% inter word space
    
    %%%%%%%%%%%%%%%%%%%%%%%%% Low temperature bonding techniques for 3DICs %%
    Currently, there exist five major low-temperature bonding techniques that address the bonding needs in Three Dimensional Integrated Circuit (3DIC) devices: direct, surface activated, eutectic, adhesive, and nano-metal bonding. Each bonding technology uses different approaches and material combinations. Some of these potential technologies are good candidates for mass production applications, and among them, the low temperature hybrid bonding by eutectic solder seems to have significant advantages due to less damage to the device. Cu-Sn system has become a popular material system for 3D integration. It overcomes the issues related to the eutectic systems (Au-ln, ln-Sn, etc.), where achieving a good performance is highly process-dependent and difficulties such as freezing may occur during reflow, leading to poor wetting and strength. It also overcomes the environmental issues related to the Pb-Sn systems enabling more than Moore system scaling in the 3DICs \cite{ko_low_2012,ohnuma1999thermodynamic,arden_more-than-moore_????}.
    
    %%%%%%%%%%%%%%%%%%%%%%%%% TLPB and affecting parameters %%
Overall, the system down-scaling induces the transition from soft sphere-shaped solders to hard cylindrical intermetallic (IMC) solders as shown in fig.~\ref{fig:domain}(a). While the soft sphere-shaped solders form thin IMC layers over Cu metallization layers, the second type (narrow interconnections) have near-bamboo grain structures due to similar line widths or narrower than the average grain diameter of the original interlayer \cite{joo_analytic_1994}. This dimensional transition also enforces an IMC-mediated packaging system where its integrity is controlled by the structure and morphology of the $Cu_6Sn_5$ and $Cu_3Sn$ IMC layers \cite{tu2011reliability}.
    
    %%%%%%%%%%%%%%%%%%%%%%%%% The impact of the electromigration %%
While thermomigration (TM) is not a major concern in conventional electronic interconnections, electromigration (EM) is known to be the main failure mechanism in Cu-based Through Silicon Via (TSV) technology \cite{chen2010electromigration,ceric2011electromigration,chao2006electromigration}, since interconnections are required to carry current densities on the order of 0.1 $\sfrac{MA}{cm^2}$ at temperatures above 100$^\circ$C. These large current densities induce significant mass transport (due to electromigration coupling), resulting in considerable changes in the interconnect microstructure during operation, and early failure of the device prominently due to the formation of voids \cite{chen2010electromigration,brown1995cluster}.     
	  
	The experimental studies by Gan et.~al \cite{gan_polarity_2005}, Orchard et.~al \cite{orchard2005electromigration}, and the recent study by Yao et~al. \cite{yao2017study} show the distinct characteristics of the EM/TM-mediated IMC growth at the anode and cathode layers and depict the morphological changes due to large mass fluxes. The analytical discussion on the impact of EM flux on the evolution of IMC layers by above studies does not distinguish the distinct nature of the interchange between the $Cu_3Sn$ and $Cu_6Sn_5$ layers. Our previous phase-field study \citep{park_2013_phase} on EM-mediated ripening treats the anode and cathode IMC layers on separate domains and neglects the complexities of the interchange between these layers. Chao et. al \cite{chao2006electromigration} and Gurov et. al \cite{gurov1981theory} analyzed the kinetics of the diffusion of the dual IMC layers under EM conditions and emphasized the complexities of the current-driven transport.
	      
	%%%%%%%%%%%%%%%%%%%%%%%%%%%% Vacancy evolution %%
    %%%%%%%%%%%%%%%%%%%%%%%%%%%% Vacancy evolution %%
    EM also induces enhanced exchange of atoms/vacancies in the materials. This quickly induces rapid migration of the atoms and possible generation of voids due to vacancy condensation in certain phases/regions of the microstructure, as it is observed in the $Cu_{3}Sn$ IMCs during service \cite{tu2011reliability,deorio2010physically}. The EM reliability issue in flip-chip systems is now known to be a design issue and increasing the Cu Under Bump Metallization (UBM) thickness to 50 $\mu$m primarily solves the current density issue by enabling better distribution of the current \cite{chen2010electromigration}. However, dimensional constraints in 3DIC technologies preclude the same strategy from being used to ameliorate the more considerable EM-induced void formation issues in microjoints. 
    
    %%%%%%%%%%%%%%%%%%%%%%%%%%%% What to do!!!! %%
    %%%%%%%%%%%%%%%%%%%%%%%%%%%% What to do!!!! %%
    A summary of the experimental microstructures of the $Cu/Sn/Cu$ sandwich interconnection is provided in Table \ref{tab:TLPB_experiments}. Apart from the experimental studies, numerous numerical models (FDR \cite{gusak_kinetic_2002}, LSW \cite{suh2008size}, phase-field \citep{attari_phase_2016,park_concurrent_2012}) have been utilized to study the growth and ripening process of IMCs during soldering and conventional transient liquid phase bonding (TLPB) processes. This work specifically covers the EM-mediated growth of individual IMCs and distinguishes the distinct nature of the interchange between the anode and cathode layers by carrying out the simulations in one domain. This addresses the influence of the top/bottom layer on each other and the difference in the rate of growth of IMCs on both sides. In addition, it enables addressing the EM-mediated vacancy transport from one side to the other by considering vacancy generation/annihilation in the structure and studying the impact of microstructureal components on vacancy transport. The work is also motivated by the necessity of establishing reliable, scale-bridging numerical methods \cite{panchal_key_2013} for studying the microstructural phenomena in the materials, in order to characterize their response under service conditions. This is essential in the case of metallic microjoints such as the ones used in 3DIC packaging systems \cite{ceric2011electromigration}. This goal makes it necessary to consider different, potentially multi-scale microstructural processes \cite{curtin2003atomistic} in the materials to gain an understanding of the underlying physical phenomena responsible for the observed responses while breaking the evolving discontinuities among these scales \cite{deborst2008challenges}. 
    
    Accordingly, we have built an integrated computational framework to incorporate the multi-phase-field method \cite{attari_phase_2016,park_concurrent_2012}, addressing the microstructural evolution in the Cu/Sn/Cu sandwich structure during complex liquid and solid state reactions with the EM and vacancy evolution models. In this way, we couple the modified phase-field approach with the electric charge continuity equation to study the evolution of the miscrostructure of the interconnection under external electric fields. In contrary to several EM studies in the literature and our previous study \citep{park_2013_phase}, we take different effective nuclear chagre values for the various components of the microstructure. Furthermore, we analyze the crystal structure of the IMCs using Density Functional Theory (DFT) method \cite{kohn1965selfconsistent} coupled with a thermodynamic framework \cite{ding2015pydii} to calculate the thermodynamics of the point defects and solute site preferences in the Cu/Sn/Cu system. Accordingly, we study the vacancy transport by considering vacancy generation/annihilation in the structure. This work utilizes the foundations of Integrated Computational Materials Engineering (ICME) methodology \cite{pineda_achieving_2014,allison_integrated_2006} to illustrate the application of different computational tools to analyze the microstructures across different scales ranging from atomic to micro scales as it is the nature of the problem in this study. 
    
    %%%%%%%%%%%%%%%%%%%%%%%%%%%%%%%%
    %%%%%%%%%%%%%%%%%%%%%%%%%%%%%%%%
    %%%%%%%%%%%%%%%%%%%%%%%%%%%%%%%%
    %%%%%%%%%%%%%%%%%%%%%%%%%%%%%%%%
  
    %%%%%%%%%%%%%%%%%%

    \newcolumntype{T}{ >{\centering\arraybackslash\LARGE} A{0.35}{0.35} }
    \newcolumntype{C}{ >{\centering\arraybackslash\LARGE} A{0.45}{0.35} }
    \newcolumntype{D}{ >{\centering\arraybackslash\huge} A{0.1}{0.1} }
    \newcolumntype{E}{ >{\centering\arraybackslash\huge} A{0.25}{0.25} }
    \newcolumntype{F}{ >{\centering\arraybackslash\huge} A{0.65}{1.10} }
    \newcolumntype{G}{ >{\centering\arraybackslash\huge} A{0.30}{0.30} }
    
    \begin{table*}[htbp]
    \caption{Summary of recent experimental studies on the formation of $Cu/Sn/Cu$ microjoints}\label{tab:TLPB_experiments}
    \centering
    \scalebox{0.35}[0.28]{

    \makebox[1.0pt]{
    \begin{tabular}{E D D G G G G G F}
    \toprule
    
    \multirow{2}{*}{} & \multirow{2}{*}{\rotatebox[origin=c]{90}{\parbox[c]{3.0cm}{\centering Reference}}} & \multirow{2}{*}{\rotatebox[origin=c]{90}{\parbox[c]{3.0cm}{\centering Process}}} & \multicolumn{2}{C}{Initial domain} & 
    \multicolumn{3}{F}{Process conditions} & \multirow{2}{*}{Selected Microstructure} \\ 
    
     &  &  & Sn layer thickness ($\mu m$) & Cu layer & Temperature ($^\circ$C) & Bonding pressure & Duration (min) & \\ \cmidrule{1-9} 
    
    \multirow{3}{*}{\rotatebox[origin=c]{90}{\parbox[c]{14.0cm}{\centering Joint Microstructure}} } & \cite{yao2017study} & \rotatebox[origin=c]{90}{TLPB} & 6 & $5\times5\times1$ & 260 & 1 N & 300 & \includegraphics[width=9cm]{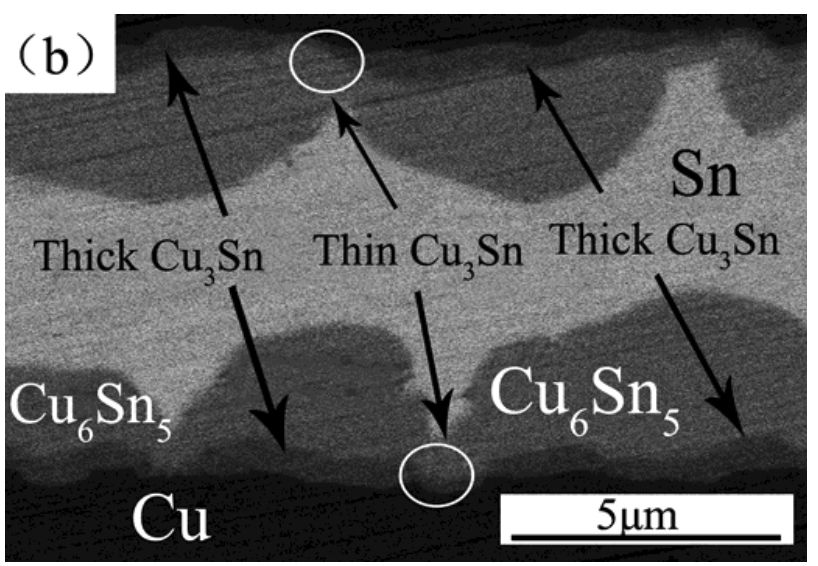} \\
    
     & \cite{chunjin_hang_phase_2013} & \rotatebox[origin=c]{90}{TLPB} & 30 & $3\times3\times1$ & 240 & 0.05 N & 960 & \includegraphics[width=9cm]{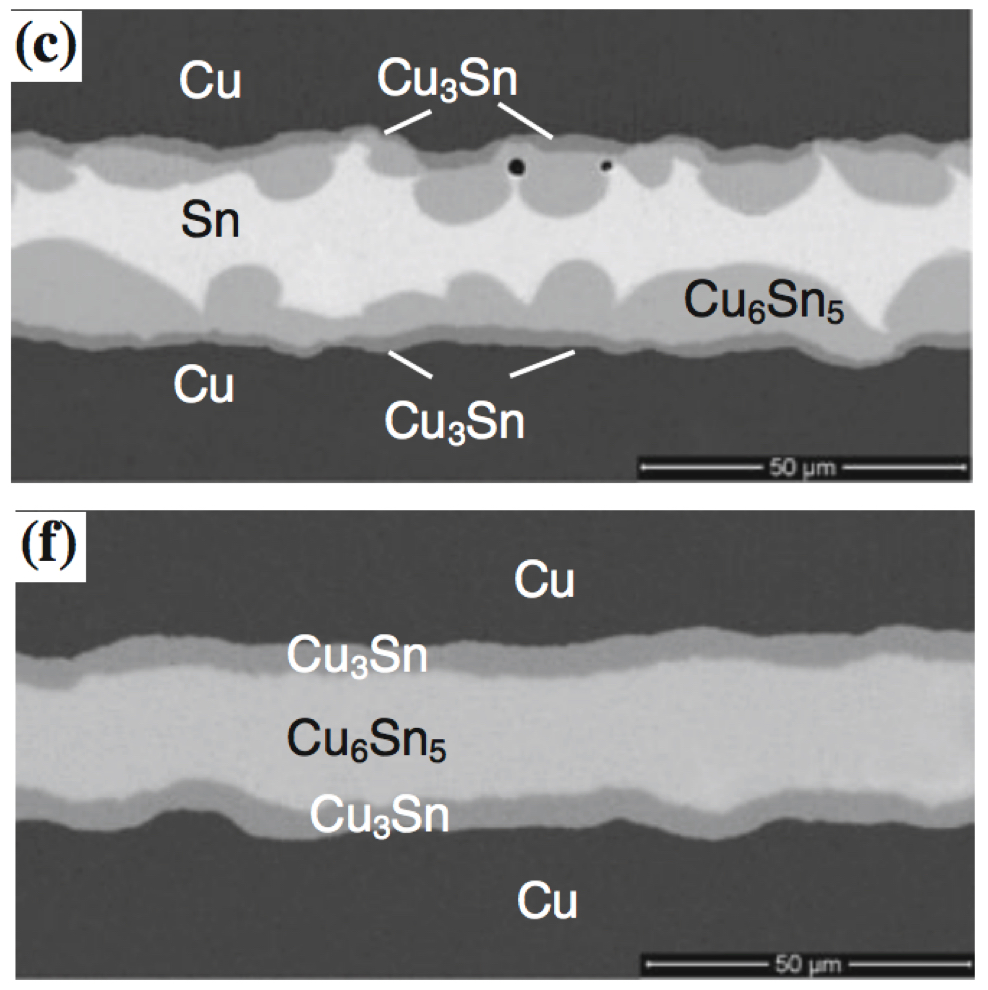} \\
    
     & \cite{zhao2017noninterfacial} & \rotatebox[origin=c]{90}{TLPB} & 20 & $5\times5\times0.5$ & 250 & 0.2 MPa & 120 & \includegraphics[width=9cm]{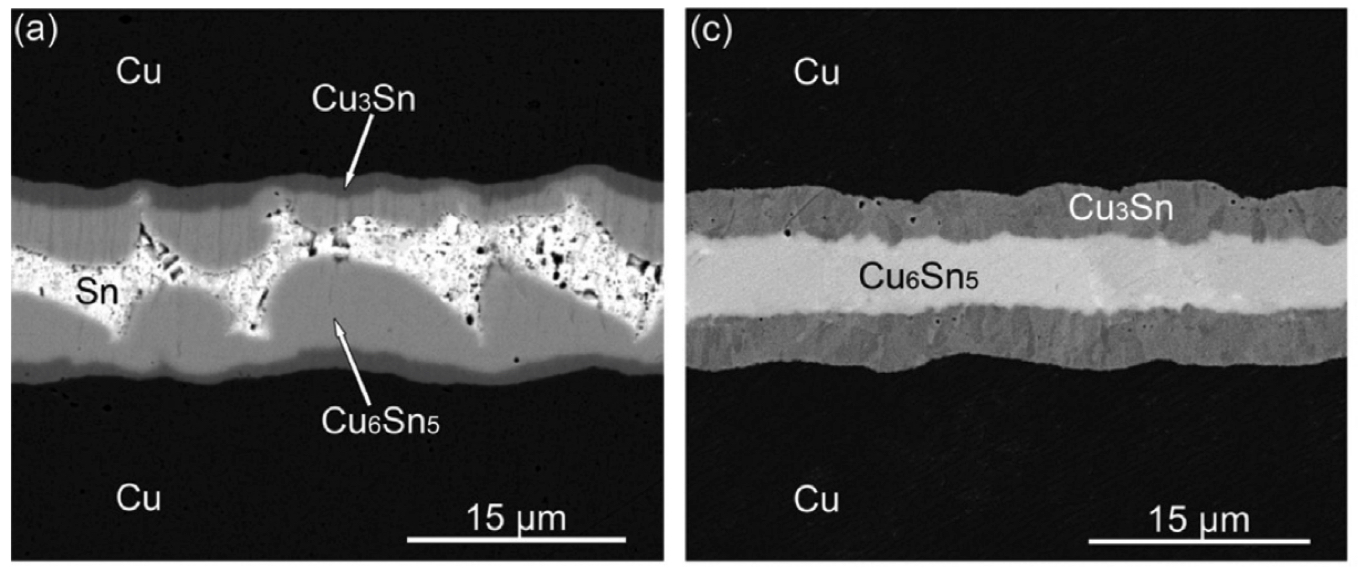} \\
    
    \multirow{2}{*}{\rotatebox[origin=c]{90}{\makecell{Joint microstructure during \\ accelerated EM/TM}}} & \cite{feng2016influence} & \rotatebox[origin=c]{90}{Bonding under EM} & 30 & $2.5\times2.5\times1$ & 260 & 0.5 N & 480 & \includegraphics[width=9cm]{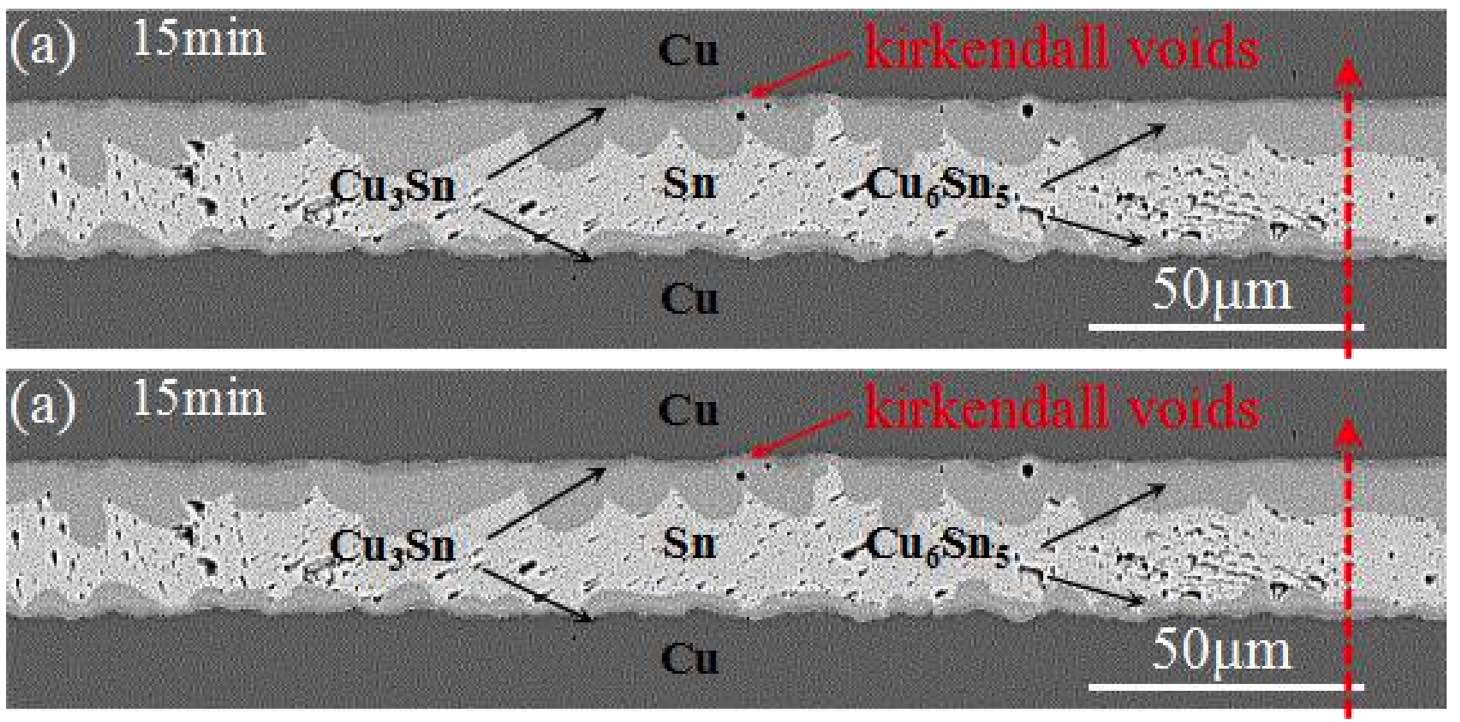} \\
    
    & \cite{yang2016full} & \rotatebox[origin=c]{90}{Bonding under TM}& 10 & $3\times3\times0.3$ & 260,300,350 & No data & 5 & \includegraphics[width=9cm]{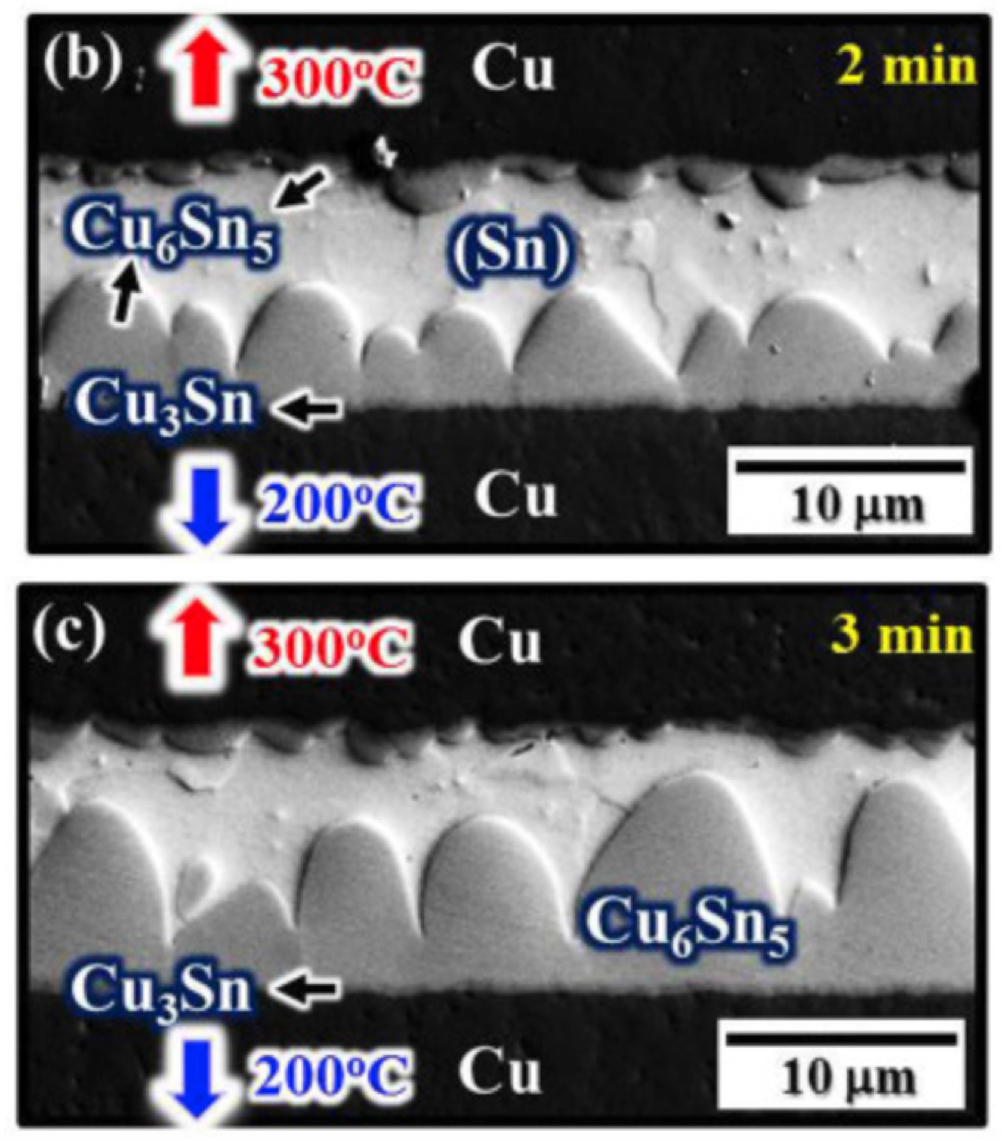} \\ \bottomrule
    
    \end{tabular}%
    }}
    \end{table*}
    
    %%%%%%%%%%%%%%%%%%

    \section{Methods}
    \subsection{Multi-Phase-Field Method} \label{sec:PFM}
    
    In this study, a multi-phase-field formalism is used to study the evolution of microstructure at isobaric and isothermal states. The components of the material system evolve based on the variational principles of total free energy of the material. A set of non-conserved ($\phi$) and conserved variables ($c$) describe the components of the microstructure, where the non-conserved variables define the spatial fraction of available phases over the domain ($\Omega$) and the conserved variables define the phase compositions. We start from a general model of total free energy of a chemically heterogenous system that involves interfacial, bulk and electrical interactions: 
    
    \begin{equation}\label{eq:fenergy_gen}
    \centering
        F^{tot} = \int_{\Omega}{F^{int} + F^{bulk} + F^{elec}} 
    \end{equation}
    
    \noindent
    where the three contributing factors are respectively formulated as: 
    
    \begin{equation}\label{eq:fenergy_gb}
    \centering
        F^{int} = \sum_{j>i}\sum_{i}\Big[\frac{\epsilon^2_{ij}}{2}\nabla \phi_{i}.\nabla \phi_{j}+\omega_{ij} \phi_{i} \phi_{j}\Big]
    \end{equation}    
    
    \begin{equation}\label{eq:fenergy_chem}
    \centering
        F^{bulk} = \sum_{i}\phi_{i}f_{i}^{0}(c_{i}) 
    \end{equation}
    
    \begin{equation}\label{eq:fenergy_elec}
    \centering
        F^{elec} = \sum_{i}\phi_{i}f_{i}^{elec}(c_{i}) = N_A e\psi\sum_{i}{Z_{i}^{*}c_i}
    \end{equation}

    \noindent 
    with $\epsilon_{ij}$ as gradient energy coefficient representing the energy penalty in interface between $i^{th}$ and $j^{th}$ phases, and $\omega_{ij}$ as a double-well potential accounting for the transformation barrier of the phase transition. The gradient energy coefficient ($\epsilon_{ij}$) and the barrier height ($\omega\textsubscript{ij}$) are functions of the interface energy ($\sigma\textsubscript{ij}$) and the interface width ($w_{ij}$). $f^{0}$ is the free energy of the homogenous system and $c_{i}$ is the molar concentration of the phases. In this study, all concentrations are based on molar concentration of Sn in each phase. $N_A$ is the Avogadro number, $e$ is the natural charge of an electron, $\psi$ is the electronic potential, and $Z^*$ is the the effective nuclear charge of the phases.
    
    Using the model of the total free energy as a function of the field variables $(\phi_i)$ and $(c)$, we postulate the following form of the kinetic equations (phase-field and diffusion) as the governing equations:

    \begin{equation}\label{eq:PFE}
    \centering
    \begin{aligned}
    \frac{\partial \phi_{i}}{\partial t} & = - \sum_{i \neq j}\frac{M_{ij}}{N_p} \bigg[\frac{\delta F^{tot}}{\delta \phi_i} - \frac{\delta F^{tot}}{\delta \phi_j} \bigg]
    \end{aligned}
    \end{equation}
    
    \begin{equation}\label{eq:diff}
        \frac{\partial c_{i}}{\partial t} = \nabla .\bigg[D(\phi_{i})\sum_{i}\phi_{i}\nabla{c_{i}}\bigg]
    \end{equation}
    
    \noindent
    where $M_{ij}$ is the interface mobility, and $N\textsubscript{p}$ is the number of coexisting phases at the neighboring grid points. The phase-field equation works only in the interface where $\phi_i$ changes between $0$ and $1$. Defining the interdiffusion parameter $D(\phi_{i}$) as a function of the phase-field order parameter in the diffusion equation (eqn. \ref{eq:diff}) allows to easily take into account the diffusivity in various features of the microstructure (i.e., interfaces, GBs, and bulk phases). 
    
\begin{figure*}[htbp]
    \vspace{-5pt}
    \centering
    \begin{center}
    \makebox[\textwidth]{\includegraphics[scale=0.055]{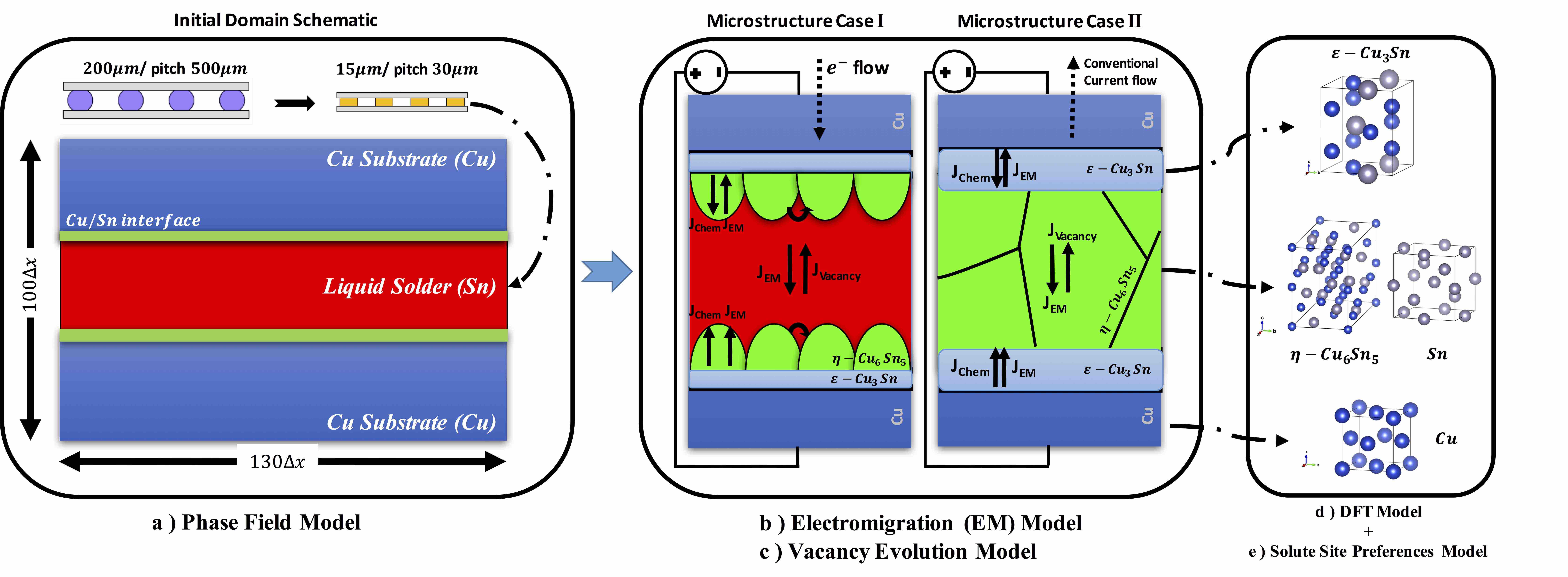}}
    \end{center}
    \vspace{-5pt}
    \caption{a) The domain schematic used in the phase-field model prior to the nucleation of the IMCs. The transition from soft solder to the IMC-mediated joint is shown in the top of (a). b) Illustration of the initial microstructures in EM and vacancy evolution studies. c) The orthorombic crystal structure of $Cu_3Sn$ and the monoclinic crystal structure of $Cu_{6}Sn_5$ used in DFT calculations (small spheres represent Cu atoms and larger spheres represent Sn atoms).}
    \label{fig:domain}
    \vspace{-10pt}
    \end{figure*}

    We solve the governing equations over a two-dimensional domain using a finite difference solver discretized in space and time. The selected domain shape and the relative dimensions are based on the recent trend in evolution of the size and shape of the solder bumps in 3DIC industry, where the thickness of the microjoint is less than 10 $\mu m$ (refer to Table \ref{tab:TLPB_experiments}). TLPB requires that the atoms in the bulk substrate dissolve into the Sn interlayer \cite{cook_overview_2011}. Consequently, a thin Cu/Sn amorphous solution, (see \cite{pan_amorphous_2008} for details) as it is observed experimentally, forms over the substrate. In this study, a molten Sn layer in the middle, and two Cu substrates at the top and bottom along with two thin (2 grid points) Cu/Sn solution layers in between the Sn and Cu sections are set as the initial condition of the material system. The initial state of the microstructural domain, prior to the nucleation of the secondary phases, is shown in fig.~\ref{fig:domain}(a). The system is supposed to adjust itself to reach a steady-state regime by nucleation and growth of the interfacial IMCs according to thermodynamics and kinetics of the system. The boundary conditions are periodic at the right and left side of this domain and Neumann with zero flux at the top and bottom. The grid points in the solution domain may comprise a mixture of different phases. The coexisting phases in the interface is stated by the condition of equality of chemical potentials and the mass conservation as \cite{huh_phase_2004}:
    
      \begin{equation}\label{eq:chemcond}
        f^{1}_{c_1}[c_{1}(x,t)] = f^{2}_{c_2}[c_{2}(x,t)] = ... = f^{i}_{c_i}[c_{i}(x,t)]
      \end{equation}
      \begin{equation}\label{eq:conserv}
        c(x,t) = \sum_{i=1}^{N} \phi_{i}c_{i}
      \end{equation} 
      
    \noindent  
    where $f^{i}_{c_{i}}$ stands as the derivative of the free energy with respect to the composition of the phase $i$. The free energies have the form of the CALPHAD functional expression for a binary system, fulfilling the thermodynamic requirements as a function of composition and temperature \cite{saunders_calphad_1998}:
    
    \footnotesize
    \begin{equation}\label{eq:cfenergy}
    f^i(c_i,T) = \sum_{i} c_i G_i^0 + RT\sum_{i}{c_i ln(c_i)} + \sum_{i} \sum_{j>i}c_i c_j \sum_{\nu} L_{ij}^{\nu} (c_i-c_j)^{\nu}
    \end{equation}
    \normalsize
    
    \noindent 
    with $R$ as the ideal gas constant, and $L_{ij}^{\nu}$ the excess binary interaction parameter which is dependent on the value of $\nu$. In this study, $\nu=0$ and the system is regular. The reference Gibbs energy term, $G_{i}^{0}$ and $L_{ij}^{\nu}$, are obtained from \cite{shim1996thermodynamic}. The system of nonlinear equations shown in eqns. \ref{eq:chemcond} and \ref{eq:conserv} are solved at each iteration of the main solver. 
    
    The instability in the pure Cu/Sn/Cu material system starts around the melting point of Sn (220$^\circ$C) where the diffusion is considerable. By increasing the process temperature, liquid viscosity and subsequently the surface tension decreases \cite{manko2001solders}. Both of these parameters assist the spreadability of the IMCs over the substrate. Hence, we investigate the evolution of the low volume solder interconnection during three common reflowing temperatures of 260$^{\circ}$C, 300$^{\circ}$C and 340$^{\circ}$C. The same simulations are repeated for different values of the $Cu_{6}Sn_{5}/Sn$ interface energy ($\sigma_{\eta L}$) to investigate its impact on the evolution of IMCs.
    
    %%%%
    \subsubsection{Nucleation of the Intermetallics} \label{sec:nuc}
    
    The phase-field method is a self-consistent field theory and an additional nucleation theory is essential to consider the nucleation of new grains. We use the classical nucleation theory with the discrete Poisson probability distribution to facilitate the likelihood of independent nucleation events at fixed intervals of time and/or space with an average occurrence rate. The approach is initially developed by Simmons et al. \cite{simmons_phase_2000} in the context of phase-field modeling to avoid the use of Langevin noise term in the classical phase-field formulation. The Poisson nucleation probability, $P_n$, and the rate of the nucleation events, $I$, in an undercooled liquid is given by
   
    \normalsize
    \begin{equation}\label{eq:7}
        P_{n} = 1- exp\bigg[-(I.v. \Delta t)\bigg]
    \end{equation} 
      \begin{equation}
        I=I_{0}exp\bigg[-\frac{16\pi\sigma^3}{3k_BT(\Delta G_{v})^2}\frac{\cos^3\theta-3\cos\theta+2}{4}\bigg]
    \end{equation} 
    \normalsize
    
    \noindent  
    with $\Delta t$ as the time interval of nucleation, \emph{v} as volume of a nucleus, and $I$ as the nucleation rate. $I_{0}$ is the nucleation rate constant (a kinetic prefactor), altered in between the suggested values on the order of ($10^{31}\pm1$ {${m^{-2}}{s^{-1}}$}) for the surface based nucleation. $\sigma$ is the interface energy, $\Delta G_{v}$ is the barrier height of the nucleation, and $\theta$ is the contact angle. The detailed study of the impact of the nucleation parameters on the microstructure can be found in \cite{attari_phase_2016}.
    
    %%%%%%%%%%%%%%%%%%%%%%%%%%%%%%%%%%%%%%%%%%%%%%%%%%%%%
    %%%%%%%%%%%%%%%%%%%%%%%%%%%%%%%%%%%%%%%%%%%%%%%%%%%%%
    
    \subsection{Solid State Growth of the IMCs During Accelerated Electromigration}
    
    The net flux of the atoms due to the chemical potential gradients ($\nabla {\mu}$) and the external electric field ($\vec{E}=-\nabla{\psi}$) is:
    
    \begin{equation}\label{eq:elflux}
    \vec{J}_{net} = \vec{J}_{chem} + \vec{J}_{em} 
    \end{equation}
    
    \begin{equation}\label{eq:EMflux}
    \vec{J}_{net} = -D(\phi_{i})\sum_{i}\phi_{i}\nabla{c_{i}} +\frac{D(\phi_{i})c_{i}}{k_{B}T} \sum_{i}{c_i}{\phi_i}.eZ^{*}(\phi_{i}) \psi
    \end{equation}
    
    \noindent
    Similar to the diffusion constant, the effective charge parameter, $Z^*$, is defined as a function of the phase-field variable at each grid point. $k_B$ is the Boltzmann constant and $T$ is operation temperature. In the context of phase-field modeling, the respective diffusion equation for studying the evolution of the microstructure under an external electric field with the constraint, $c= \sum_i c_i\phi_i$ is:
    
    \begin{equation}\label{eq:EM}
    \frac{\partial c}{\partial t} = \nabla . \bigg[D(\phi_{i})\sum_{i}\phi_{i}\nabla{c_{i}} - \frac{D(\phi_{i})}{k_{B}T} \sum_{i}{c_i}{\phi_i}.eZ^{*}(\phi_{i}) {\psi}\bigg]
    \end{equation}
    
    Equation \ref{eq:EM} is coupled with the phase-field equation (eqn. \ref{eq:PFE}) for modeling the solid state growth of the IMCs under the influence of external electric field. 
    
    The initial microstructure is selected from the phase-field simulations when no electrical flow is present. Accordingly, the isothermally reflowed virtual microstructures at elevated temperatures of 260$^\circ$C, 300$^\circ$C and 340$^\circ$C undergo accelerated EM conditions at 130$^\circ$C, 150$^\circ$C and 180$^\circ$C reaction temperatures.
    
    We assume that the applied electric field and the solid media are quasi-static and isotropic, respectively. Consequently, the distribution of the electrostatic potential over the domain is calculated assuming the quasi-stationary conduction process. In this way, the continuity equation ($\frac{\partial \rho}{\partial t} + \nabla.\vec{J_e}=\dot{\pi}$) for the charge conduction simplifies to $\nabla.{\vec{J_e}}=0$. Hence, the equation below (also called Ohm's equation) is used to obtain the electrostatic potential over the domain:
    
    \begin{equation} \label{eq:ohm}
    \nabla .\bigg{[{\kappa(c_i,\phi_i)}{\nabla {\psi}}\bigg]} = 0
    \end{equation}
    
    \noindent
    where $\kappa$ is the conductivity of the material. Equation \ref{eq:ohm} is solved separately during each iteration of the phase-field solver by using the following boundary and initial conditions:
        
        \small
        \begin{equation}
          \arraycolsep=1.4pt\def\arraystretch{1.0}
          \left\{
                    \begin{array}{ll}
                      \frac{\partial \psi}{\partial x}\bigm\vert_{top} \:\;\;\;\; = -J_e.\kappa(c_{i},\phi_{i})\\
                      \frac{\partial \psi}{\partial x}\bigm\vert_{bottom} = -J_e.\kappa(c_{i},\phi_{i})\\
                      \text{Periodic at left and right sides. }\\
                      \psi\bigm\vert_{t=0} \quad \;\;\;= 0.95 \quad (V)\\
                    \end{array}
           \right.            
        \end{equation}
        \normalsize
    
    This initial condition is the common working potential for the Xilinx 20nm 3DIC (UltraScale series) according to \cite{leibson2013xilinx}. Consequently, the electrostatic field, electric field, current density distribution and mass flux and their vector gradients along with chemical evolution of the material system are calculated. Additionally, EM induces a minor change in the equilibrium concentration values. Hence, the equilibrium concentration values under the influence of the electrical fields are re-calculated by construction of the electro-chemical potentials due to the addition of $\sum_{i} f^{i}_{em}(c_i)$ term to the total free energy.
    
    %%%%%%%%%%%%%%%%%%%%%%%%%%%%%%%%%%%%%%%%%%%%%%%%%%%%%
    %%%%%%%%%%%%%%%%%%%%%%%%%%%%%%%%%%%%%%%%%%%%%%%%%%%%%
    
    \subsection{Vacancy Transport During Electromigration}
    
    Point defects influence many physical properties and have role in many diffusion controlled processes. Due to the lack of experimental studies ( e.g., positron annihilation spectroscopy, lattice parameter or electrical resistivity measurements) on the characterization of vacancy formation and migration in the Cu/Sn/Cu system, we use a DFT approach \cite{kohn1965selfconsistent} in combination with a dilute solution thermodynamic formalism \cite{ding2015pydii} in order to determine the equilibrium properties of the point defects and the solute site preferences in IMCs. Later, based on the obtained information, we employ a vacancy transport model to analyze the transient evolution of vacancies during the evolution of the microstructures.

    %%%%%%%%%%%%%%%%%%%%%%%%%%%%%%%%%%%%%%%%%%%%%%%%%%%%%
    %%%%%%%%%%%%%%%%%%%%%%%%%%%%%%%%%%%%%%%%%%%%%%%%%%%%%
    
    \subsubsection{DFT Calculation and the Equilibrium Concentration of the Point Defects}
    
    DFT \cite{kohn1965selfconsistent} framework, as implemented in the Vienna ab-initio simulation package (VASP) \cite{kresse1996efficient,kresse1996efficiency} is used to calculate the energy spectrum in the IMCs. Generalized gradient approximation (GGA) is used as the DFT functional as it is proposed by Perdew, Burke, and Ernzerhof \cite{perdew1996generalized}. The GGA appears to be more reliable in Sn-based IMCs \cite{lee2006structural}. The electronic configurations of the elements in the system (Cu and Sn) are determined by the projector augmented-wave (PAW) \cite{blochl1994projector}. Brillouin zone integrations were performed using a Monkhorst-Pack mesh \cite{h.j2005special} with at least 3000 k points per reciprocal atom. Full relaxations were realized by using the Methfessel-Paxton smearing method \cite{methfessel1989highprecision} of order one and a final self-consistent static calculation with the tetrahedron smearing method with Bl{\"o}chl corrections \cite{blochl1994improved}. A cutoff energy of 533 eV was set for all calculations and the spin polarizations were taken into account. The relaxations were carried out in three stages: first stage by allowing changes in shape and volume, corresponding to the (VASP) ISIF = 7 tag—, second stage by additionally allowing the relaxation of atoms, corresponding to the (VASP) ISIF = 2 tag and a final self-consistent static calculation run. The equilibrium concentration of the vacancies and their formation energies are investigated by the grand-canonical, dilute-solution thermodynamic formalism proposed in Ref.~\cite{ding2015pydii}.
    
    %%%%%%%%%%%%%%%%%%%%%%%%%%%%%%%%%%%%%%%%%%%%%%%%%%%%%
    %%%%%%%%%%%%%%%%%%%%%%%%%%%%%%%%%%%%%%%%%%%%%%%%%%%%%
    
    \subsubsection{Transient Evolution of Vacancies in the System}
    
    For a sufficiently dilute system where mass transport involves diffusion via vacancies, it is convenient to introduce vacancy as an additional component to balance the transport of the moving atoms with respect to a reference system fixed to the average atomic velocity: 
    
    \begin{equation}
        J_v = - \sum_{i}{J_i} \qquad \text{( i=1,2, \dotso ) }
    \end{equation}
    
    \noindent
    where $i$ stands as the number of the distinct atoms in the material system, and the vacancies are supposed to be in thermal equilibrium everywhere. Large electric fields or thermal gradients are necessary to impose considerable deviation of the vacancy concentration from equilibrium. In addition, the electrochemical flux of the vacancies is combined with a vacancy sink term accounting for the rate of the change in vacancy concentration due to diffusion to a fixed number of sinks (annihilation). It is already known that point defects can be produced/annihilated in excess of the thermodynamic equilibrium vacancies \cite{damask1963point}. There exists a thermodynamic driving force to cut the concentration of the defects to the equilibrium concentration of the crystal. Excess defects may disappear from a crystal by two different mechanisms, migration to the sinks or recombination. Hence, the rate at which vacancies can accumulate at particular points in the sandwich interconnection is expressed as:
    
    \begin{equation}\label{eq:vac1}
    \frac{\partial c_v}{\partial t} = \nabla. \vec{J}_{v} - (c_v-c_{eq}(\phi_i))\times (\frac{\Gamma}{n})
    \end{equation}
    
    \noindent
    where $\vec{J}_{v}$ is the vacancy flux due to the chemical and electrical interactions. The second term is the rate of change of vacancies due to diffusion to a fixed number of sinks when vacancy concentration is out of the equilibrium. $c_v$ is the vacancy concentration in the sample at a certain time $t$ out of equilibrium vacancy concentration $c_{eq}$. $\Gamma$ is the jumping frequency of the vacancy at temperature $T$ and $n$ is the average number of jumps needed for a vacancy to overtake to reach a sink. $\Gamma$ is given by $\Gamma=\nu \times exp\big(-\frac{E_M}{k_B T}\big)$ where $\nu$ is the average effective vibration frequency and $E_M$ is the vacancy migration energy. Substitution into eqn. \ref{eq:vac1} results into the following partial differential equation for the evolution of the vacancies:
    
    \small
    \begin{equation}\label{eq:vac2}
    \begin{aligned}
    \frac{\partial c_v}{\partial t} ={} & \nabla. \Bigg[D_v^{1}(\phi_{i})\nabla{c_{v}} - \frac{D_v^{2}(\phi_{i})}{k_{B}T} {c_v}.eZ^{*}(\phi_i) \psi\Bigg] - {\frac{\nu}{n}} \big(c_{v}-c_{eq}(\phi_{i})\big) \times exp\big(\frac{-E_M(\phi_i)}{{k_B}T}\big) 
    \end{aligned}
    \end{equation}
    
    \normalsize
    
    \noindent
    where $D_v$ is the true statistical vacancy diffusion coefficient under EM condions, which are related to the intrinsic diffusion constants via a correlation factor ($f$). Here, we avoid the discussion on the difficulty of obtaining the value of $f$. In a dilute binary system, the chemical diffusion of vacancies ($D_v^1$) can be approximated by $D_{Sn}$-$D_{Cu}$ and the electrical diffusivity of vacancies ($D_v^2$) can be approximated by $D_{Sn}$+$D_{Cu}$ \cite{chao2006electromigration}. $D_{Sn}$ and $D_{Cu}$ are the intrinsic diffusivities of Cu and Sn in the phases and the values are shown in Table. \ref{tab:intrinsic_dif}.

    %%%%%%%%%%%%%%%%%%
    %%%%%%%%%%%%%%%%%%
    
    \newcolumntype{I}{ >{\centering\arraybackslash\huge} A{0.45}{0.15} }
    \newcolumntype{L}{ >{\centering\arraybackslash\LARGE} A{0.48}{0.48} }

    \begin{table}[htbp]
    \caption{Intrinsic diffusion constants used in the vacancy transport model \cite{bondy1971diffusion,kumar2011intrinsic}.}\label{tab:intrinsic_dif}
    \centering
    \scalebox{0.40}[0.40]{
    \makebox[1.0pt]{    
    \begin{tabular}{I L L L L}
    \toprule
    \diaghead{\theadfont Diag ColumnmnHead II}%
  {Element}{Phase}  & Cu                                           & $Cu_3Sn$                                     & $Cu_6Sn_5$                                  & Sn \\ 
    Cu & 2.95$\times10^{-5}e^{\frac{-43.89}{RT}}$ & 1.80$\times10^{-8}e^{\frac{-78.8}{RT}}$ & 6.20$\times10^{-8}e^{\frac{-80.5}{RT}}$  & 2.4$\times10^{-11}e^{\frac{-33.02}{RT}}$  \\
    Sn & 3.40$\times10^{-5}e^{\frac{-43.89}{RT}}$ & 7.90$\times10^{-10}e^{\frac{-79.9}{RT}}$ & 5.80$\times10^{-7}e^{\frac{-85.2}{RT}}$ & 1.2$\times10^{-9}e^{\frac{-43.89}{RT}}$  \\
    \bottomrule
    \end{tabular}
    }
    }
    \end{table}
    
    %%%%%%%%%%%%%%%%%%
    %%%%%%%%%%%%%%%%%%

    %%%%%%%%%%%%%%%%%%%%%%%%%%%%%%%%%%%%%%%%%%%%%%%%%%%%%
    %%%%%%%%%%%%%%%%%%%%%%%%%%%%%%%%%%%%%%%%%%%%%%%%%%%%%
    
    \section{Results and Discussion}
    
    We studied the formation of a Cu/Sn/Cu microjoint in the 3DIC packaging systems suitable for manufacturing conditions. This step forms the necessary initial microstructures to be used for studying the impact of EM and the subsequent morphological changes in the microjoint which are explained in section \ref{sec:elec_res}. The crystal structure of the IMCs are investigated in terms of the formation of equilibrium intrinsic point defects in section \ref{sec:vac_res}. Next, the migration of vacancies in the microstructure due to the applied external field is studied. In each section, an overview of the subsequent computational observations is provided, and the key comparisons with the experiments are pointed out.

    \subsection{Formation and Morphology of the TLPB joint (Interface energy impact)} \label{sec:formation}

    %% where do the parameters come from?
    In this section, we explore the rate and the extent of growth of the IMCs by changing the temperature and the model parameters, specifically the IMC-liquid ($\sigma_{\eta L}$) interface energy. Temperature variation subsequently affects the system thermodynamics and diffusion parameters (Arrhenius diffusivity relations are used). Interfacial mobilities are defined as a function of the diffusivity parameters. However, interfacial energies are temperature invariant constant values which are defined based on our previous knowledge in this system. We used similar values as estimated by our previous studies for larger size solder systems. The assumption by \cite{gagliano_nucleation_2002} for $\sigma_{\eta L}$ interface energy is one order of magnitude smaller than what we already used in our previous phase-field modeling study \citep{attari_phase_2016}. The Arrhenius diffusivity relations, interfacial mobilities, interfacial energies and other material properties used in our simulations are summarized in Table \ref{tab:parameters}.

%%%%%%%%%%%%%%%
%%%%%%%%%%%%%%%

\begin{table*}[htbp]

\caption{The material parameters for different components of the microstructure. \label{tab:parameters}}
\begin{center}

\scalebox{0.70}[0.75]{
\makebox[1.0pt]{

\begin{tabular}{@{}lcccccccccccc@{}}
\toprule \textbf{Microstructure Components} & \textbf{Type} & \multicolumn{9}{c}{\textbf{Material Properties}} \\ \midrule
\multirow{4}{*}{\textbf{Bulk phases}} & Sn         & \multirow{9}{*}{{\rotatebox[origin=c]{90}{\textbf{Diffusivity ($\sfrac{m^2}{s}$)}}}} & $2.47\times10^{-11} exp(-\frac{33020}{k_BT})$ & \multirow{9}{*}{{\rotatebox[origin=c]{90}{\textbf{Mobility ($\sfrac{m^2}{s}$)}}}} & n/a & \multirow{9}{*}{{\rotatebox[origin=c]{90}{\textbf{Interface Energy ($\sfrac{J}{m^2}$)}}}} & n/a & \multirow{9}{*}{{\rotatebox[origin=c]{90}{\textbf{Effective Charge ($Z^*$)}}}} & -17.97 & \multirow{9}{*}{{\rotatebox[origin=c]{90}{\textbf{Resistivity (\sfrac{1}{$\kappa$}) ($\Omega.m$)}}}} & $1.10\times10^{-7}$ \\
                                  & Cu             &  & $5.90\times10^{-5} exp(-\frac{138.8}{k_BT})$ &  & n/a &  & n/a &  & -9.10 &  & $1.70\times10^{-8}$ \\
                                  & $Cu_{6}Sn_{5}$ &  & $2.58\times10^{-8} exp(-\frac{85200}{k_BT})$ &  & n/a &  & n/a &  & -13.93 &  & $1.75\times10^{-7}$ \\
                                  & $Cu_{3}Sn$     &  & $3.65\times10^{-10}exp(-\frac{79700}{k_BT})$ &  & n/a &  & n/a &  & -15.75 &  & $9.83\times10^{-8}$ \\
\multirow{3}{*}{\textbf{Interfaces}}       & $\sfrac{Sn}{Cu_{6}Sn_{5}} (\sigma_{\eta L})$        &  & $0.2\times D_{Sn}$           &  & $(1.0\times10^6) D_{Sn}$ &  & 0.1, 0.125, 0.15 &  & -8.98 & & $2.775\times10^{-7}$ \\ 
                                  & $\sfrac{Cu_{6}Sn_{5}}{Cu_{3}Sn}$  &  & $1.5\times D_{Cu_{6}Sn_{5}}$ & & $(7.0\times10^3) D_{Sn}$ &  & 0.5              &  & -6.96 & & $2.495\times10^{-7}$  \\
                                  & $\sfrac{Cu_{3}Sn}{Cu}$ &  & $1.5\times D_{Cu_{3}Sn}$   &  & $(7.0\times10^3) D_{Sn}$              &  & 0.5       &  & -7.87 & & $1.763\times10^{-7}$ \\  
\multirow{2}{*}{\textbf{Grain Boundaries}} & ${Cu_{6}Sn_{5}}$  &  & $8\times D_{Cu_{6}Sn_{5}}$ &  & $(7.0\times10^3) D_{Sn}$ &  & 0.3       &  & -2.78  &  & $2.697\times10^{-7}$ \\ 
                                  & ${Cu_{3}Sn}$       &  & $10\times D_{Cu_{3}Sn}$ &  & $(7.0\times10^3) D_{Sn}$  &  & 0.3 &  & -3.15 & & $2.184\times10^{-7}$ \\  \bottomrule
\end{tabular}

}}

\end{center}
\end{table*}

%%%%%%%%%%%%%%%%%%%%%%%%% 
%% Initial stage (L+S) and 2nd stage (S)  %%

\begin{figure*}[h!]
\centering
\makebox[\textwidth]{\includegraphics[scale=0.06]{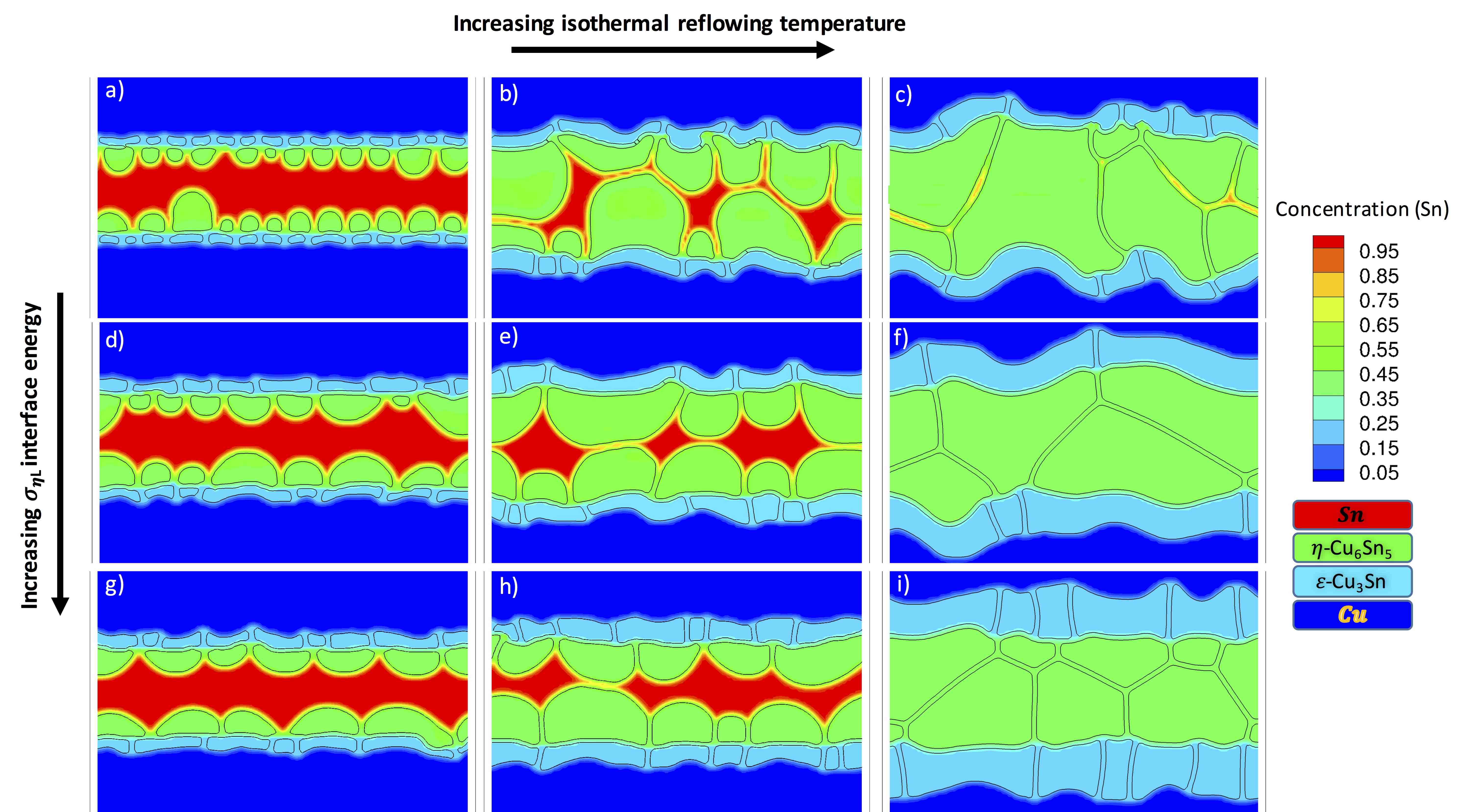}}
\caption{The microstructure of the Cu/Sn/Cu sandwich interconnection after reflowing at (a,d,g) 260$^\circ$C (b,e,h) 300$^\circ$C and (c,f,i) 340$^\circ$C for 25 minutes. The corresponding interfacial energies are (a,b,c)=0.1 $\sfrac{J}{m^2}$ (d,e,f)=0.125 $\sfrac{J}{m^2}$ and (g,h,i)=0.15 $\sfrac{J}{m^2}$. A video clip is available online. Supplementary material related to this article can be found online at (***).}
\label{fig:evolution}
\vspace{-10pt}
\end{figure*}

\begin{figure*}[htbp]
\centering
\begin{center}
\makebox[\textwidth]{\includegraphics[scale=0.04]{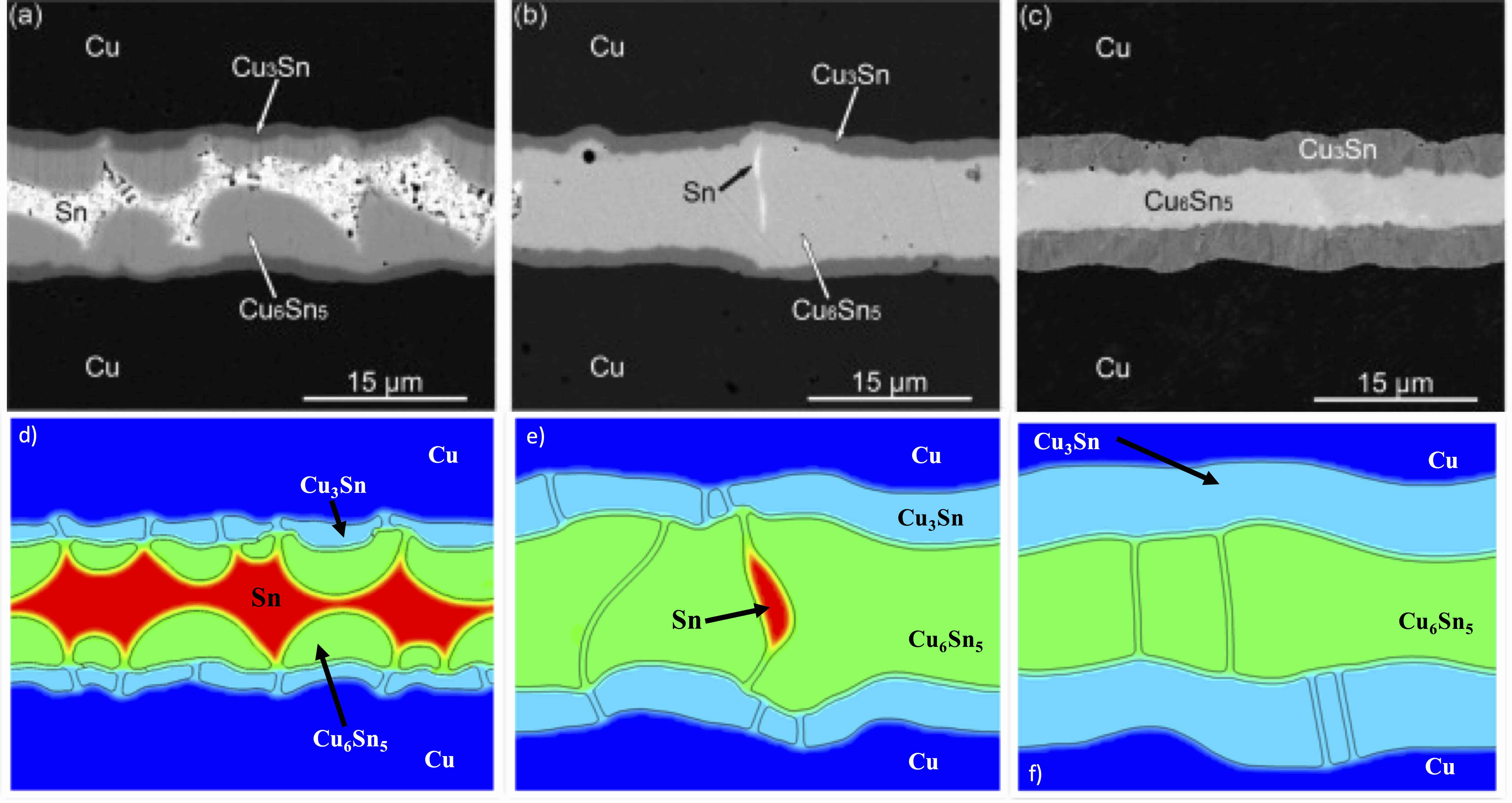}}
\end{center}
\caption{Comparing the simulations with experiments \cite{zhao2017noninterfacial} in terms of the kinetics of evolution and obtained morphology. TLPB experimental results at 250$^\circ$C after a) 20 b) 40 and c) 60 minutes. phase-field simulations at 260$^\circ$C after d) 20 e) 40 and f) 60 minutes.}
\vspace{-10pt}
\label{fig:simvsexp}
\end{figure*}

Figure~\ref{fig:evolution} demonstrates the isothermal growth of the IMCs after reflowing for 25 minutes at 260$^\circ$C, 300$^\circ$C and 340$^\circ$C. Similar to our previous TLPB joint formation study, the simulations follow two stages, the first stage of the evolution contains both liquid and solid state reactions while the second state entirely consists of solid state reactions. Each reaction type implies different equilibrium conditions, which will be satisfied by solving a set of non-linear equations, subjected to the equal chemical potential (eqn. \ref{eq:chemcond}) and mass conservation (eqn. \ref{eq:conserv}) constraints at the interface. The initial $Cu_{6}Sn_{5}$ and $Cu_{3}Sn$ nuclei are imposed uniformly in the system in a square shape by means of the nucleation module. These phases grow in scallop and planar morphologies, respectively. During the initial stages of the evolution, the $Cu_{6}Sn_{5}$ IMCs grow fast into the Sn and consume the Sn layer entirely while the $Cu_{3}Sn$ layer thickness remains almost intact. Later, equiaxed islands of the $Cu_{6}Sn_{5}$ grains form when the two $Cu_{6}Sn_{5}$ layers meet with each other entirely. At this point, the reaction changes due to full consumption of the Sn interlayer and the system enters the sequence of solid state reactions. The growth rate is considerably slow for this stage. Later, the $Cu_{3}Sn$ grains dominate the material system by consuming $Cu_{6}Sn_{5}$ grains and covering the entire interlayer. Compositions of phases in different interfaces and the reaction exponent (n) for average thickness of the IMC layers during liquid stage reactions for different simulations are reported in Table~\ref{tab:reac_cst}. The study shows that the reaction rate is independent of solder volume or substrate area fraction and it is more dependent on the amount of bulk and interfacial diffusive features in the microstructure. We avoid providing further details of the reaction process during TLPB here, and readers may refer to our previous work \cite{attari_phase_2016}. An example morphological comparison between the current phase-field study and the experimental study of Cu/Sn/Cu sandwich interconnection with 15 $\mu m$ interlayer thickness is illustrated in fig.~\ref{fig:simvsexp}. 

%% Morphology of both IMCs at the initial stage
We have already shown that the diffusivity values, particularly along the interfaces and GBs affect the growth rate and morphology of the IMCs rather than the amount of the solder material \citep{attari_phase_2016}. However, the interfacial energy variation has not been addressed before. Whether the effect of $\sigma_{\eta L}$ interface energy and $Cu_6Sn_5$ GB diffusivity rate are dependent or independent requires sensitivity analysis. Both these parameters promote the growth of $Cu_6Sn_5$ IMCs, and their growth rate has significant impact on the morphology of this IMC. Hence, based on our previous studies on determination of working parameters for the phase-field model and also with current set of Arrhenius diffusivities, $\sigma_{\eta L}$ interface energy value should be roughly 0.1$< \sigma_{\eta L}<$0.15. The value of the interface energy in between the $Cu_{3}Sn$ IMCs should always be 2 to 3 times higher compared to that of the $Cu_{6}Sn_{5}$ IMCs, in order to obtain a planar morphology of the $Cu_{3}Sn$ layer.

%%%%%%%%%%%%%%%%%%

%%%%%%%%%%%%%%%%%%

Figure~\ref{fig:evolution} depicts how increasing the value of $\sigma_{\eta L}$ parameter from 0.1 to 0.15 $\sfrac{J}{m^2}$ decreases the wetting angle. While the nucleus starts as a square in tangent contact with the substrate, its height then rises as the capillary wave initiated from the triple line arrives at the top of the nucleus. Contact angle hysteresis is observed in the phase-field modeling as it is often seen in practice. Oscillations may occur in the triple-line position and the largest contact angle oscillation is roughly close to 1.5 radian for the case of 0.1 $\sfrac{J}{m^2}$. The $Cu_{6}Sn_{5}$ IMCs grow with bamboo-like morphologies when the interface energy is 0.1 $\sfrac{J}{m^2}$. This behaviour is very similar to the effect of the $Cu_{6}Sn_{5}$ GB diffusion where IMCs grow faster into the liquid with higher rates, although an increase in $Cu_{6}Sn_{5}$ GB diffusivity value never yields bamboo-shaped structure.

%% When and how coalescence occurs? (Cu6Sn5)
%% The liq-Cu6 interface energy impact
The variation of $\sigma_{\eta L}$ interface energy points out the question that when and how the coalescence in $Cu_{6}Sn_{5}$ IMCs occurs. As experiments also suggest, careful tracing of the coalescence of $Cu_{6}Sn_{5}$ IMCs with the neighboring grains show that an IMC should be at least twice the size of the neighboring grain in order to absorb it. From a simulation point of view, this only happens when the $\sigma_{\eta L}$ interface energy is higher than 0.125 $\sfrac{J}{m^2}$. On the contrary, there is no coalescence when $\sigma_{\eta L}=$ 0.1 $\sfrac{J}{m^2}$. Hence, we hypothesize the effect of dynamic interface energy on the evolution of $Cu_{6}Sn_{5}$ IMCs. This can be confirmed by looking at the experimental observations in fig.~1 of the study by Choi et al. \cite{choi2000effect}. The micrographs in this figure clearly show that the interface energy should be around 0.1 $\sfrac{J}{m^2}$ upto 120 seconds of the reaction time, while for elongated reaction times the value seems to be more on the side of (0.125 to 0.15) $\sfrac{J}{m^2}$. In the later cases, the coalescence continues while the $Cu_{6}Sn_{5}$ top and bottom layers touch each other. At this point, the formed islands of IMCs change the growth regime in favor of the touching $Cu_{6}Sn_{5}$ phases and ultimate formation of a single $Cu_{6}Sn_{5}$ layer. On the other hand, computations indicate that the penetration of the Sn liquid into the $Cu_{6}Sn_{5}$ triple junctions increases by increasing the value of $\sigma_{\eta L}$ from (0.1 to 0.15) $\sfrac{J}{m^2}$. 

%%%%%%%%%%%%%%%%%%
%%%%%%%%%%%%%%%%%%

%%%%%%%%%%%%%%%%%%
%%%%%%%%%%%%%%%%%%

%% When and how coalescence occurs in (Cu3Sn) layer?
The $Cu_{3}Sn$ IMCs also merge with each other, while maintaining the planar layer morphology during the initial liquid state period. When the top and bottom $Cu_{6}Sn_{5}$ layers fully touch, the extent of growth of the $Cu_{3}Sn$ layer alters. Consequently, the $Cu_{3}Sn$ gets dominant in the microstructure by consuming the entire $Cu_{6}Sn_{5}$ IMCs as it is shown in fig.~\ref{fig:simvsexp}. Significant calibration of the parameters is necessary to capture the exact behaviour in the $Cu_{3}Sn$ IMC layer. Also, we have not observed any quantitative morphological changes due to the variation of the temperature. This is mainly due to the application of temperature invariant interfacial energies for the various components of the microstructures.

%%%%%%%%%%%%%%%%%%
%%% 
%%% 

%There is a considerable conflict in determining the dominant diffusive species during the extended solid state reactions. $Cu$ seems to be the dominant diffusing specie through the GBs and Sn has to pass through the bulk IMC grains. Energy Dispersive Spectrometer (EDS) spot analysis indicate arterial Cu flux to the sides of the $Cu_6Sn_5$ IMC from the substrate \cite{mu2016critical}. In that case, Sn should mostly be adsorbed from $Cu_{6}Sn_{5}$ IMCs instead of the grain boundary channels. This may conclude that the $Cu_3Sn$ IMC layer has kinetically little chance to grow comparing to the $Cu_6Sn_5$ layer. However, based on the thermodynamics and the observations, it should be there in between the Cu UBM and the $Cu_{6}Sn_{5}$ IMCs to maintain the local equilibrium conditions across the interface.

\begin{table*}
\normalsize
\renewcommand{\arraystretch}{1.00}
\centering
\caption{The summary of calculated equilibrium concentration in the interfaces, and EM free reaction exponents for the $Cu_6Sn_5$ IMCs growth rate at different temperatures and $\sigma_{\eta L}$ energy values.}
\label{tab:reac_cst}
\resizebox{\linewidth}{!}{%
\begin{tabular}{@{}ccccccccccc@{}}
\toprule
\multicolumn{2}{c}{\textbf{Reflowing condition}} & \multicolumn{8}{c}{\textbf{Concentration in the interface of each phase (Molar Sn)}} & \textbf{Reaction exponent} \\ \midrule
Temp ($^\circ$C) & $\sigma_{\eta L}$ ($\sfrac{J}{m^2}$) & $C_{(Cu/Sn)}^{Cu}$ & $C_{(Cu/Sn)}^{Sn}$ & $C_{(Cu_{6}Sn_{5}/Sn)}^{Cu_{6}Sn_{5}}$ & $C_{(Cu_{6}Sn_{5}/Sn)}^{Sn}$ & $C_{(Cu_{3}Sn/Cu_{6}Sn_{5})}^{Cu_{3}Sn}$ & $C_{(Cu_{3}Sn/Cu_{6}Sn_{5})}^{Cu_{6}Sn_5}$ & $C_{Cu/Cu_{3}Sn}^{Cu}$ & $C_{Cu/Cu_{3}Sn}^{Cu_{3}Sn}$ & n \\
\multirow{3}{*}{260} & 0.1 & \multirow{3}{*}{0.562} & \multirow{3}{*}{0.958} & \multirow{3}{*}{0.439} & \multirow{3}{*}{0.973} & \multirow{3}{*}{0.247} & \multirow{3}{*}{0.433} & \multirow{3}{*}{0.0287} & \multirow{3}{*}{0.226} & 0.307 \\ 
 & 0.125 &  &  &  &  &  &  &  &  & 0.282 \\
 & 0.15 &  &  &  &  &  &  &  &  & 0.221 \\
\multirow{3}{*}{300} & 0.1 & \multirow{3}{*}{0.550} & \multirow{3}{*}{0.941} & \multirow{3}{*}{0.435} & \multirow{3}{*}{0.958} & \multirow{3}{*}{0.246} & \multirow{3}{*}{0.432} & \multirow{3}{*}{0.034} & \multirow{3}{*}{0.226} & 0.393 \\
 & 0.125 &  &  &  &  &  &  &  &  & 0.186 \\
 & 0.15 &  &  &  &  &  &  &  &  & 0.194 \\
\multirow{3}{*}{340} & 0.1 & \multirow{3}{*}{0.537} & \multirow{3}{*}{0.920} & \multirow{3}{*}{0.434} & \multirow{3}{*}{0.937} & \multirow{3}{*}{0.246} & \multirow{3}{*}{0.432} & \multirow{3}{*}{0.039} & \multirow{3}{*}{0.226} & 0.493 \\
 & 0.125 &  &  &  &  &  &  &  &  & 0.192 \\
 & 0.15 &  &  &  &  &  &  &  &  & 0.372 \\ \cmidrule(lr){1-11}   
\end{tabular}
}
%\vspace{0pt}
\end{table*}

\normalsize

%%%%%%%%%%%%%%%%%%%%%%%%%%%%%%%%%%%%%%%%%%%%%%%%%%%%%%%
%%%%%%%%%%%%%%%%%%%%%%%%%%%%%%%%%%%%%%%%%%%%%%%%%%%%%%%
%%%%%%%%%%%%%%%%%%%%%%%%%%%%%%%%%%%%%%%%%%%%%%%%%%%%%%%

\subsection{Microstructure Evolution During Accelerated EM}\label{sec:elec_res} 

%% Drift velocity in Cu(Sn)
It is experimentally shown that EM damage in Cu(Sn) alloys increases non-linearly as a function of the Sn content in the Cu matrix. In addition, the drift velocity measurements in Cu alloys show that Sn drastically reduces the Cu migration rate \cite{hu1996electromigration}. The wind force and the subsequent current stressing enhance diffusion and the subsequent reaction kinetics in the direction of electric field, resulting in a new dominant kinetic mechanism. In other words, EM changes the IMC growth trend in the multi-layered interconnection systems as observed in the study of Gan et al. \cite{gan_polarity_2005}. This study consistently showed that under (4$\times10^7 \sfrac{A}{m^2}$) or higher current densities, the IMCs in the anode side grow faster than the IMCs in the cathode side. The growth of anode is amplified further with increasing current density. Similarly, the study by Hu et al. \cite{hu2003electromigration} shows the extensive dissolution of Cu in the cathode layer and failure of the flip-chip bump only after 95 minutes, under 2.5$\times 10^8$ $\sfrac{A}{m^2}$, while Joule heating effect also caused a temperature increase up to 157$^\circ$C at the backside of the chip. This dissolution of Cu atoms from the IMCs in the cathode layer has been confirmed to occur at a average rate of about 1 $\sfrac{\mu m}{min}$ \cite{hu2003electromigration}.

\begin{figure*}[htbp]
\centering
\makebox[\textwidth]{\includegraphics[scale=0.062]{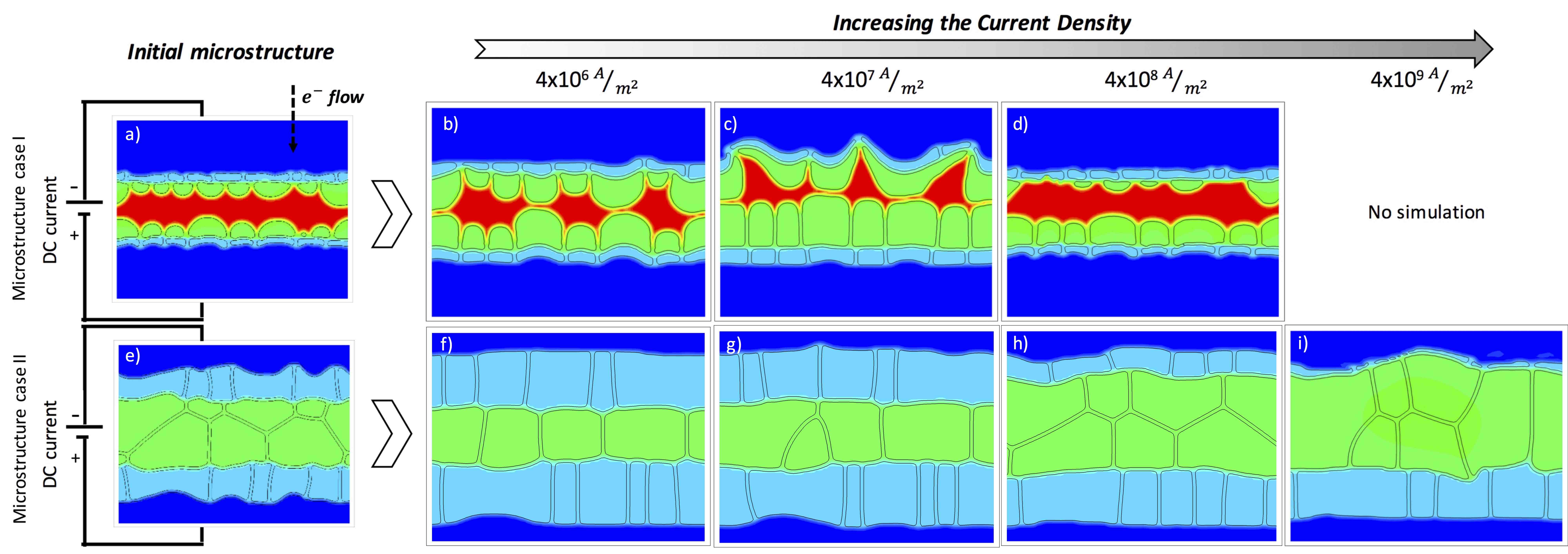}}%
\caption{The resistance of the two microstructures against Cu dissolution under different current densities at 180$^\circ$C. Top row (Case 1 or Sn-retained microstructure) from left to right: a) Initial state, b) $4\times10^{6}$ \sfrac{A}{$m^2$} after 125 hours, c) $4\times10^{7}$ \sfrac{A}{$m^2$} after 8 hours, d) $4\times10^{8}$ \sfrac{A}{$m^2$} after 20 minutes. Bottom row (case 2 or Sn-exhausted microstructure) from left to right: e) Initial state, f) $4\times10^{6}$ \sfrac{A}{$m^2$} after 80 hours, g) $4\times10^{7}$ \sfrac{A}{$m^2$} after 80 hours, h) $4\times10^{8}$ \sfrac{A}{$m^2$} after 80 hours, i) $4\times10^{9}$ \sfrac{A}{$m^2$} after 36 hours.}\label{fig:elec2}
\vspace{-10pt}
\end{figure*}

A similar trend in growth of the IMC layers was also confirmed computationally in \cite{park_2013_phase} on the impact of EM on IMC evolution. However, the work was not successful in addressing the dissolution of Cu at the cathode side as it is observed experimentally. Unlike that work, the whole anode and cathode layers are integrated in one simulation domain in this paper. This enables direct comparison and realization of the interactions of the top and bottom layers and further study of vacancy evolution when current passes through the joint. Such a realization can be observed in the first row of fig.~\ref{fig:elec2}, where under 4$\times 10^7$ $\sfrac{A}{m^2}$ current density, the shape of the IMCs is affected by the complicated nature of current routes from the bottom layer to the top side. In the next paragraphs, we will elaborate the performed computational experiments and the observed behaviour of the IMC layers.

Considering the above fact about the effect of Sn amount on EM damage, two types of initial conditions (microstructures) are selected for studying the evolution of the IMCs under the accelerated EM conditions. A microstructure consisting of all phases, including the unreacted Sn interlayer as the first case study and a Sn-exhausted microstructure fully composed of the IMCs as the second case, is selected. In the first case, the thermodynamic state of the unreacted Sn phase is changed to the solid state. The first case consists of the grains in the form of scallops with perpendicular GBs to the electric field; and in the second case, the equiaxed grains with low and high angle GBs are present in the interlayer. The second case is also the one often observed in 3DIC microjoints due to small dimensions and multiple reflows \cite{ko_low_2012}. To reveal the effect of electric field on the TLPB joints, different current densities were employed by increasing the temperature from 130 to 180$^\circ$C. In particular, the evolution of the IMC layers are investigated under $4\times10^6$, $4\times10^7$, $4\times10^8$ and $4\times10^9$ $\sfrac{A}{m^2}$ current densities. Among the several results on the morphology of the cathode and anode IMC layers according to the current density and temperature variations, we report four significant observations of morphological evolution at 180$^\circ$C for all current densities in fig.~\ref{fig:elec2}. The obtained morphologies at 130$^\circ$C and 150$^\circ$C are similar to the case of 180$^\circ$C. %Fig.~\ref{fig:elec2} depicts the obtained results for such initial microstructures.

%% intro (the impact of the Z* in the electromigration)
In this section, the diffusivities are based on the same Arrhenius relations shown in Table \ref{tab:parameters}. We assume that the main contribution to the EM flux is due to the effective nuclear charge (Z*) of different components of the microstructure, rather than the change in the value of diffusivity. Moreover, the early simulations on detecting the impact of diffusivity/interface energies did not produce a noticeable change in the results. On the other side, Z* variation showed apparent contributions in the driving force for EM induced migration and subsequent morphological changes. The range of proposed Z* values to some extent is wide in the literature. In the case of Sn, the example values are $\frac{-80}{f}$ \cite{kharkov1960electric} and $\frac{-18}{f}\pm$2 \cite{khosla1975electromigration} where $f$ is the estimated correlation factor $(f\approx0.5)$ , while in the case of Cu, the example values are $-4.8\pm1.5$ and $-5.5\pm1.5$. The values of Z* in GBs and bulk phases usually do not differ by more than one order of magnitude \cite{ho1989electromigration}. In this work, unlike our previous study \citep{park_2013_phase} where all components of the microstructure were assumed to have a constant Z* value; Z* is described as a function of the phase composition $(c)$ and the phase fraction $(\phi)$ in the continuum space by means of the Clementi-Raimondi rule \cite{clementi1963atomic}. In the pure phases and the IMCs, the values are calculated from the linear combination of the constituents of each phase. The Z* and resistivity values of the main components of the microstructure are reported in Table \ref{tab:parameters}.

%% Initial Mic: all phases are present
The first row of fig.~\ref{fig:elec2} illustrates the growth pattern of the Sn-retained microstructure under different current density conditions at 180$^\circ$. In this case, the evolution proceeds as usual by applying $4\times10^6$ $\sfrac{A}{m^2}$ current density to the joint. The $Cu_{6}Sn_{5}$ IMCs at the cathode and the anode side evolve into the Sn interlayer with almost similar growth rates as if there was no applied electric field. When $4\times10^7$ $\sfrac{A}{m^2}$ current density is applied, the evolution is considerably influenced by the electric field due to the imposed momentum on the atoms. The thickness of $Cu_{6}Sn_{5}$ IMC layer at the cathode side diminished and the growth rate significantly reduced, while the same IMCs at the anode side started to grow faster by consuming Sn. When $4\times10^8$ $\sfrac{A}{m^2}$ is applied, the $Cu_{6}Sn_{5}$ IMCs at the cathode side dissolved rapidly in the Sn, while the same IMCs at the anode side grew faster than the previous case. This is believed to be due to the release of Cu atoms from the IMCs at the cathode layer and the subsequent migration toward the anode layer. Accordingly, the separate treatment of the anode and cathode layers by Park et al.~\cite{park_2013_phase} ignores the impact of the critical interlayer length \cite{gan_polarity_2005,tu2003recent} on causing EM damage in shorter interconnections.

%% microstructure with Cu3Sn and Cu6Sn5
On the other hand, the Sn-exhausted microstructure (second row of fig.~\ref{fig:elec2}) was intact under $4\times10^6$ and $4\times10^7$ $\sfrac{A}{m^2}$ current densities and no EM-induced change was observed. The $Cu_3Sn$ IMCs at the anode layer started to grow faster by applying a current density of $4\times10^8$ $\sfrac{A}{m^2}$ or higher at the expense of $Cu_6Sn_5$ IMCs. At the same time, the $Cu_{3}Sn$ IMCs at the cathode side shrinked. Under $4\times10^8$ $\sfrac{A}{m^2}$ current density, the rate of overall growth of IMCs is very low compared to the earlier case where Sn phase is still present in the domain. While $4\times10^8$ $\sfrac{A}{m^2}$ current density acts destructively by causing $Cu_{6}Sn_{5}$ IMCs dissolve rapidly in the first case, the dissolution is slow in the second microstructural case. Since the rate of evolution of the microstructure is low in the second case compared to the Sn-retained microstructure, we also apply $4\times10^9$ $\sfrac{A}{m^2}$ in Sn-exhausted microstructural case. Accordingly, the results show that the overall rate of dissolution of Cu from $Cu_3Sn$ IMC layer is fairly low compared to Sn-retained microsture. The studies show that the service life of the second microstructural case under accelerated EM conditions could be higher than the case where Sn phase is present. Although one should also take into account the impact of vacancy migration and void formation under EM conditions.

%%%%%%%%%%%%%%%%%%%%%%%%%

%%%%%%%%%%%%%%%%%%%%%%%%%

\begin{figure}[!htbp]
\centering
%\makebox[\textwidth]{\includegraphics[scale=0.045]{flux_mass_fig.jpg}}
\includegraphics[scale=0.045]{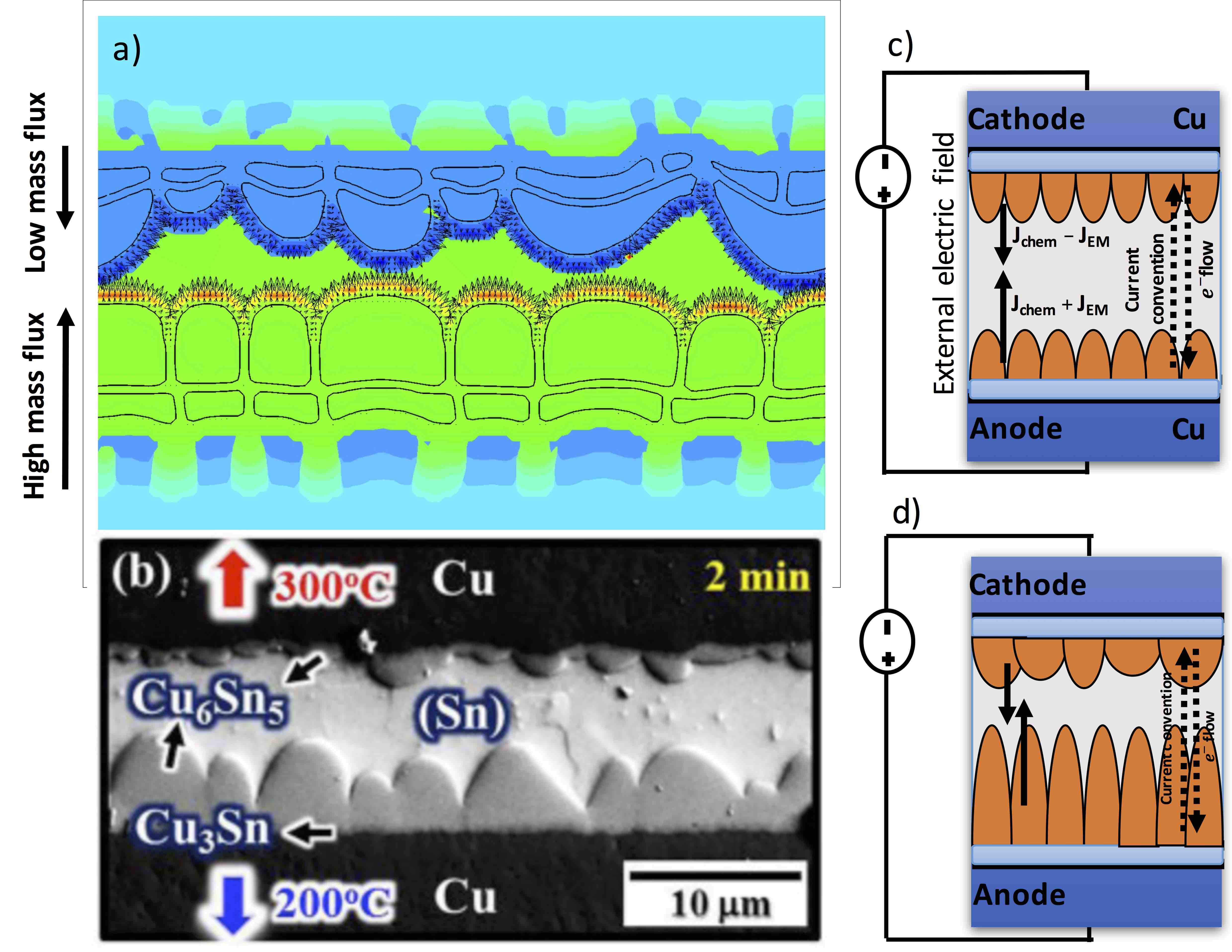}
\caption{ a) Microstructure case I: High upward mass flux in the bottom IMC layer and low downward mass flux in the top IMC layer during the early stages of the EM under $4\times10^7 \sfrac{A}{m^2}$ current density at  180$^\circ$C. b) Similar trend in growth of IMC layers observed in 6$\mu m$ Cu/Sn/Cu interconnection during TM \cite{yang2016full}. c) and d) schematic of rival fluxes during EM conditions.} \label{fig:elec1}   
\vspace{-10pt}
\end{figure}

Figure.~\ref{fig:elec1}(a) depicts the rate of mass flux in the Sn-retained microstructure under EM conditions. The locations where the mass flux is at pick are also identified by quivers; and the direction and size of the quivers imply the rate of this growth at the specific points in the IMC layers. A close look at the flux quivers reveals a high mass flux upward in the $Sn/Cu_{6}Sn_{5}$ interface (anode side) compared to the slightly lower, downward flux at the same interface in the top layer (cathode side). This result infers faster growth of the $Cu_{6}Sn_{5}$ IMCs at the anode and the corresponding shrinkage at the cathode side. This behavior is due to the competition between the chemical and electrical fluxes at the cathode versus the cooperation of these fluxes in the anode layer. These results differentiate the impact of different diffusion routes in the microstructure, indicating that the interfaces which are horizontal to the current direction induce the highest changes in the microstructure by confronting the flux flow. Figure~\ref{fig:elec1}(b) illustrates a similar trend of evolution during TM condition \cite{yang2016full} where the IMCs in one side grow faster than the other side. The physical nature of EM and TM and their impacts are known to be similar. The schematics in fig.~\ref{fig:elec1}(c) and fig.~\ref{fig:elec1}(d) illustrate such a trend in the evolution of IMCs under EM/TM conditions.

The flux curves along a vertical line (line(*)) in the microstructures are illustrated in fig.~\ref{fig:uniform_vac}(e) and \ref{fig:uniform_vac}(f). In the case where Sn phase is still present, there is a considerablse mass transport in the $Sn/Cu_{6}Sn_5$ interface, inducing considerable change in the microstructure. The simple analogy of the flux through ideal GBs versus the bulk structure under EM circumstances in the polycrystalline materials also suggests abnormal mass flux through GBs, specially at the elevated temperatures $(T>0.5T_m)$. The quivers shown in fig.~\ref{fig:elec1} also show that mass flux divergences occur at the intersection between the Sn phase and $Cu_{6}Sn_5$ IMCs where the transition between the fast and slow diffusing components of the microstructure occurs. In the second microstructual case, the flux is very high in the $Cu_3Sn/Cu_6Sn_5$ interface (fig.~\ref{fig:uniform_vac}(f)). Also, the GBs of the $Cu_{6}Sn_{5}$ IMCs evolve in the direction of the applied electric field to fulfill the tendency of the microstructure in reducing the amount of interface and reaching a steady state condition \cite{brown1995cluster}.

%% Comparing with the literature %%
%% Gan et al approach
The computational experiments in this work show that the response of each of the $Cu_3Sn$ and $Cu_6Sn_5$ IMC layers under electric field is dramatically different than the previous work, and one IMC type may resist the current crowding more than the other. This can be confirmed by checking the microstructures in fig.~\ref{fig:elec2}. For example, the $Cu_3Sn$ IMC layer is still stable while $Cu_6Sn_5$ dissolves in Sn in the case of $4\times10^7 \sfrac{A}{m^2}$ current density. Under the EM conditions, the IMC growth exponent deviates from the parabolic growth regime $(n\approx\sfrac{1}{2})$ to a linear growth regime $(n\approx1)$ at 4$\times 10^8$ $\sfrac{A}{m^2}$ current density. This infers that the growth of the overall $Cu_3Sn$ and $Cu_6Sn_5$ anode IMC layer changes from diffusion controlled regime to the chemical reaction-dominant mechanism, indicating that the mass transfer from/to the anode IMC layer is controlled not by the long-range diffusion but by the rate at which the reactants are put into contact.

%%%%%%%%
%%%%%%%%
%%%%%%%%
%%%%%%%%
%%%%%%%%
%%%%%%%%

%%%%%%%%%%%%%%%%%%%%%%%%%%%%%%%%%%%%%%%%%%%%%%%%%%%%%%%
%%%%%%%%%%%%%%%%%%%%%%%%%%%%%%%%%%%%%%%%%%%%%%%%%%%%%%%
%%%%%%%%%%%%%%%%%%%%%%%%%%%%%%%%%%%%%%%%%%%%%%%%%%%%%%%
\newcommand*{\x}{\mathsf{x}\mskip1mu}

\subsection{Vacancy generation in the layers} \label{sec:vac_res}
\subsubsection{Intrinsic point defect concentrations}

% Vacancy in Metals and PAE experiments
In this section, the equilibrium characteristics of the point defects in the crystal structure of the bulk phases in the sandwich interconnection is discussed. The equilibrium features of the point defects of pure metals are well addressed in the literature. In particular, the vacancy formation energy ($E_F^{Cu}=0.85$-$1.1$ $eV$) and equilibrium concentration ($C_{eq}^{Cu}=$\textsc{(1.9$\pm$0.5)$\times10^{-4}$} at 1075$^\circ$C) of vacancies in Cu are calculated experimentally and computationally \cite{simmons1963measurement}. It is difficult to measure these information for Sn, due to non-evident response of pure Sn to positron trapping \cite{siegel1978vacancy}. However, the dilatometry results show that the vacancy concentration is low ( $<$3$\times10^{-5}$) even close to melting temperature \cite{balzer1979equilibrium}. 

To the best of authors' knowledge, the equilibrium properties of point defects in the IMCs present in this study are rarely addressed in the literature. Hence, we study the thermodyanmic/crystallographic aspects of the equilibrium point defects in the IMCs. The $C2/c$-$Cu_{6}Sn_{5}$ monoclinic and $Pmmn$-$Cu_{3}Sn$ orthorhombic crystal structures \cite{jain2013commentary}, as shown in fig.~\ref{fig:domain}(c) are used to study the behaviour of the point defects in these systems. The $Cu_{6}Sn_5$ IMCs transform to $\eta'$ monoclinic super lattice from high temperature $\eta$ hexagonal close-packed superlattice at 186$^{\circ}$C, and the transition is an order-disorder phase transformation. Both of the $Cu_{6}Sn_5$ and $Cu_{3}Sn$ IMCs have narrow composition ranges. The $Cu_{3}Sn$ crystal structure has two distinct Cu lattice sites and one Sn site in the primitive cell. In the IMCs that have relatively close packed crystal structures, the dominant intrinsic point defects are expected to be substitutional anti-sites and the vacancies. Different sublattices may have different concentration of defects, making the diffusion process complex. Four different defect types are assumed in the crystal structure of the IMCs: vacancy in Cu sites (V\textsubscript{Cu}), vacancy in Sn sites (V\textsubscript{Sn}), Cu antisite in Sn lattice points (Sn\textsubscript{Cu}), and Sn antisite in Cu lattice points (Cu\textsubscript{Sn}), and the formation energy and equilibrium concentration of these defects at 130$^\circ$C, 150$^\circ$C and 180$^\circ$ are calculated using the PyDII framework \cite{ding2015pydii}. This framework is based on the DFT approach and it has already been applied to a range of IMCs such as L1\textsubscript{2} Al\textsubscript{3}Sc \cite{woodward2001density}, NiAl \cite{ding2015pydii}, etc. to study the energy of constitutional and thermal point defects with promising results.

\begin{figure}[htbp]
    \centering
    \subfloat[]{\includegraphics[scale=0.110]{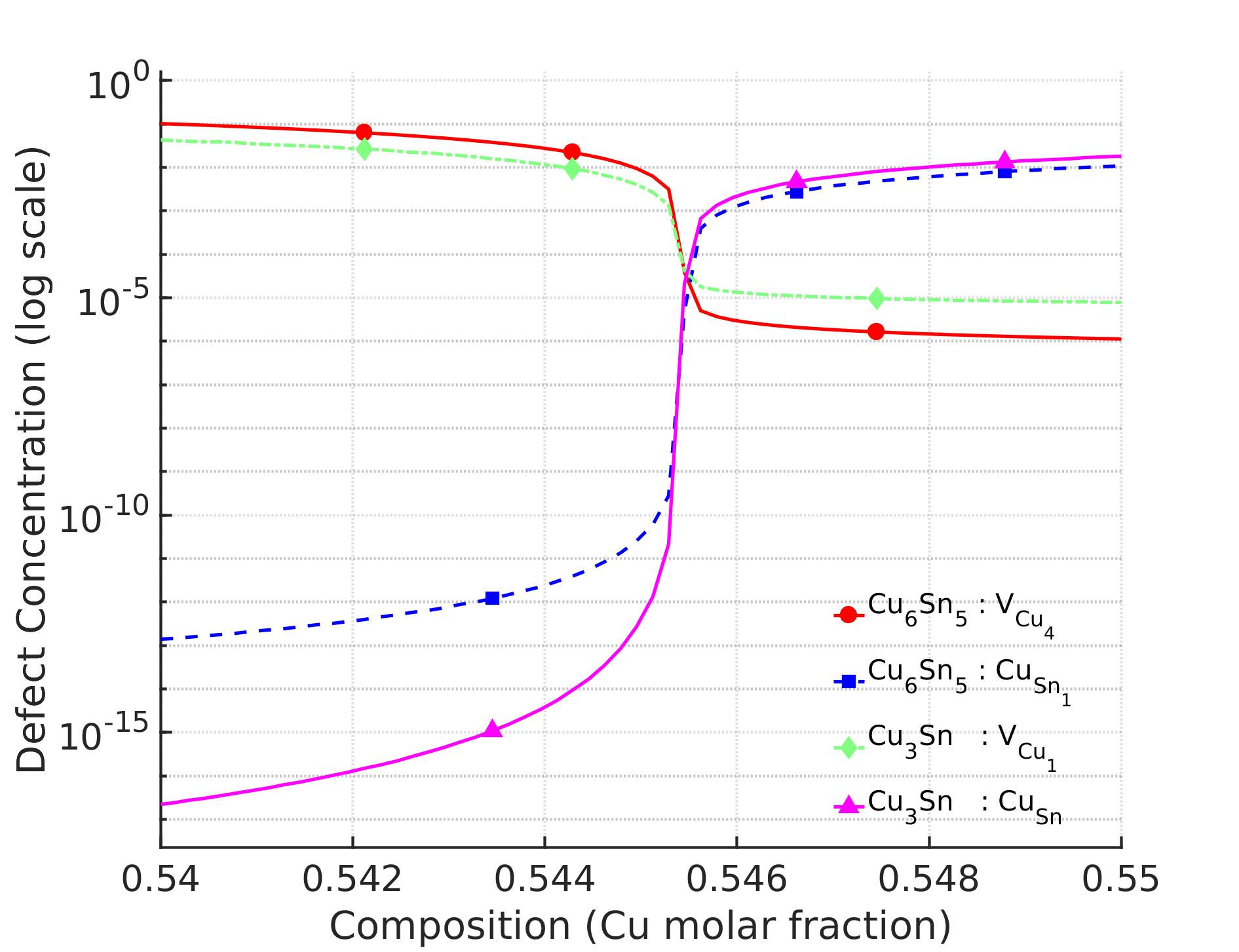}}\\
    \subfloat[]{\includegraphics[scale=0.110]{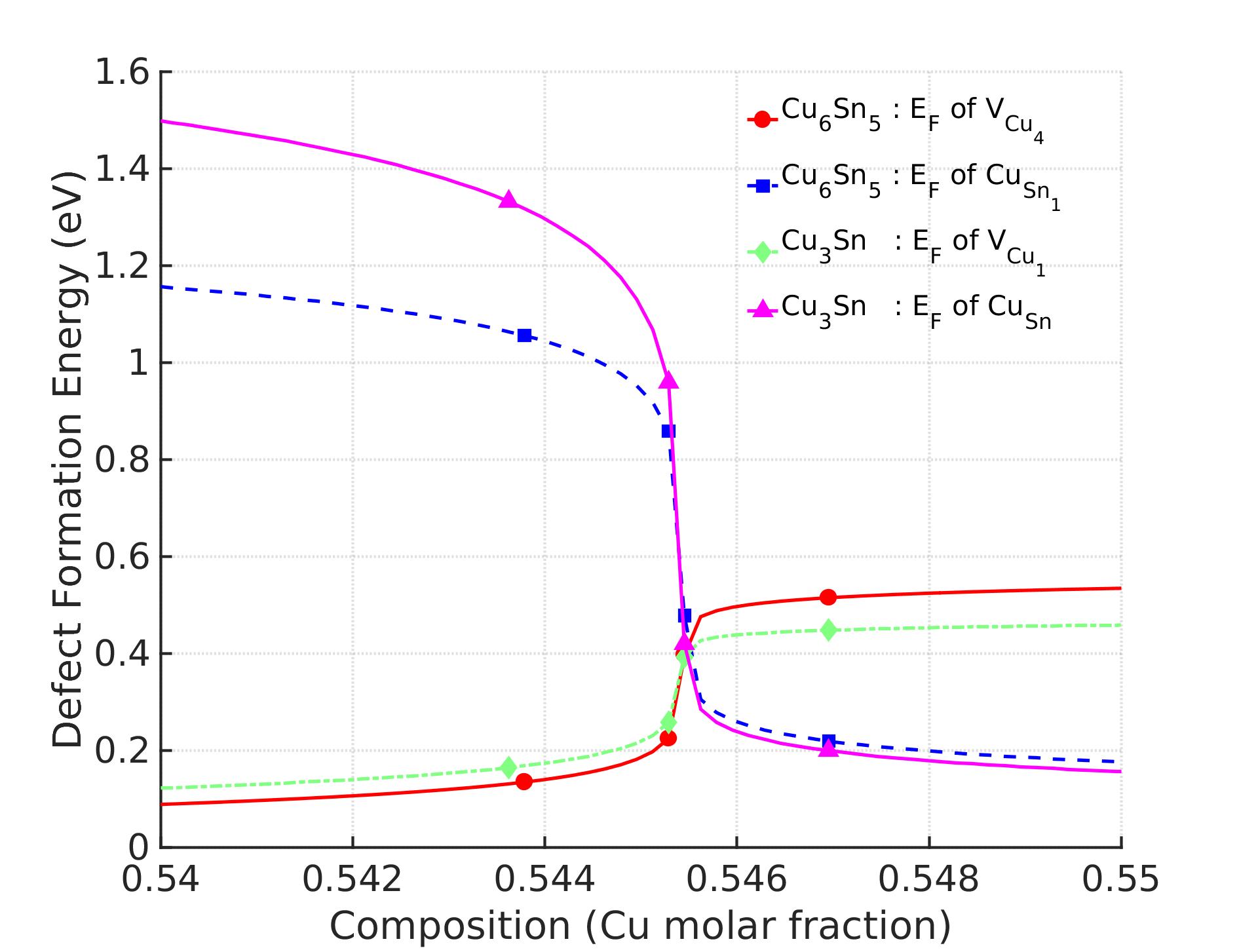}}
    \caption{a) The point defect equilibrium concentration and b) formation energy (E\textsubscript{F}) in selected lattice sites of $Cu_3Sn$ and $Cu_{6}Sn_5$ IMCs at 180$^\circ$C}
    \label{fig:sub1} 
    \end{figure}

%% crystal structure  
The equilibrium vacancy concentration profiles for $Cu_{3}Sn$ and $Cu_{6}Sn_{5}$, at 180$^\circ$C, around the ideal stoichiometric concentrations of these IMCs are illustrated in fig.~\ref{fig:sub1}. The point defect concentration values at 130$^\circ$C and 150$^\circ$C follow exactly the same trend in this figure, except the fact that the amount of equilibrium point defect concentration is slightly lower at lower temperatures. Hence, we limit the discussions to the results obtained for 180$^\circ$C. At 180$^\circ$C, V\textsubscript{Cu} value in $Cu_{3}Sn$ IMC increases by three orders of magnitude up to $\approx0.1$ as the alloy goes from Cu to Sn rich. This trend is very similar for $Cu_6Sn_5$ IMC. The value of V\textsubscript{Sn} in both IMCs is considerably low. Hence, we restrict the discussions to V\textsubscript{Cu}. The calculations show that the dominant point defect for both $Cu_{3}Sn$ and $Cu_{6}Sn_{5}$ IMCs for Cu-rich concentrations is Cu\textsubscript{Sn} antisite, while for Sn-rich side is V\textsubscript{Cu}. Depending on the structure, the excess Sn preferentially sits at a specific lattice site. 
 
In both IMCs, the vacancy defects are present in Sn-rich side, while the amount of antisite defects are 12 to 15 order of magnitude lower. Under EM conditions, the chance to produce $V_{Cu}$ sites seems to be higher than $V_{Sn}$ due to easier migration of Cu atoms. There is no way the crystal structure can sustain the saturated cluster of Cu vacancies by either creating more antisite defects or point defect annihilation mechanisms. The results in fig.~\ref{fig:sub1} also infer that under normal circumstances the amount of antisite defect is significantly low in $Cu_{3}Sn$ where the Kirkendall voids are always forming. 

    %%%%%%%%%%
    %%%%%%%%%%
    %%%%%%%%%%
    
    %%%%%%%%%%
    %%%%%%%%%%
    %%%%%%%%%%
    
 % EF of point defects in the IMCs
 The formation energies of the point defects in the IMCs are shown in fig.~\ref{fig:sub1}(b) and confirm the observed concentration trend in the IMCs. The energetic comparisons also confirm that the chance of formation of high amounts of vacancies in Cu sites are higher. Accordingly, this can also get amplified under the external electric fields due to the enforced electron forces. After all, the formation of the vacancy point defects in the Sn sites involve a relatively high relaxation in the lattice structure. Hence, the calculations support the experimental observation of the formation of voids in the $Cu_{3}Sn$ IMCs and we can conclude that the $Cu_3Sn$ IMCs are vulnerable to the formation of vacancies and the subsequent voids.

%%%%%%%%%%%%%%%%%%%%%%%%%%%%%%%%%%%%%%%%%%%%%%%%%%%%%%%
%%%%%%%%%%%%%%%%%%%%%%%%%%%%%%%%%%%%%%%%%%%%%%%%%%%%%%%
%%%%%%%%%%%%%%%%%%%%%%%%%%%%%%%%%%%%%%%%%%%%%%%%%%%%%%%

\subsubsection{Transient Evolution of the Vacancies Prior to Void Nucleation}

The equilibrium point defect concentration calculations in the previous section paves the path for understanding the characteristics of the EM-mediated vacancy migration in the sandwich interconnection system. Similar to the work of Rosenberg-Ohring \cite{rosenberg1971void}, the vacancy transport is determined by the vacancy concentration gradients due to the chemical interactions (Fick\textquotesingle s second law) and the drift due to the EM induced forces when the metal is under a direct current. The vacancy sinks/sources are taken into account by means of the thermodynamic formulation for vacancy evolution shown in equation \ref{eq:vac2}. The value of the second term in eqns. \ref{eq:vac1} or \ref{eq:vac2} is negative when $c_v > c_{eq}$, meaning vacancy concentration decreases. In certain cases, when a microstructural feature acts as vacancy generator, $c_v < c_{eq}$. This can be compared to the $\gamma$ term in Korhonen et. al \cite{korhonen1993stress} study describing the vacancy recombination in dislocations or the ad-hoc term in the Rosenberg-Ohring model \cite{rosenberg1971void}. None of the earlier models consider the impact of moving boundaries in the microstructure. 

\begin{figure*}[htbp]
\centering
\includegraphics[scale=0.063]{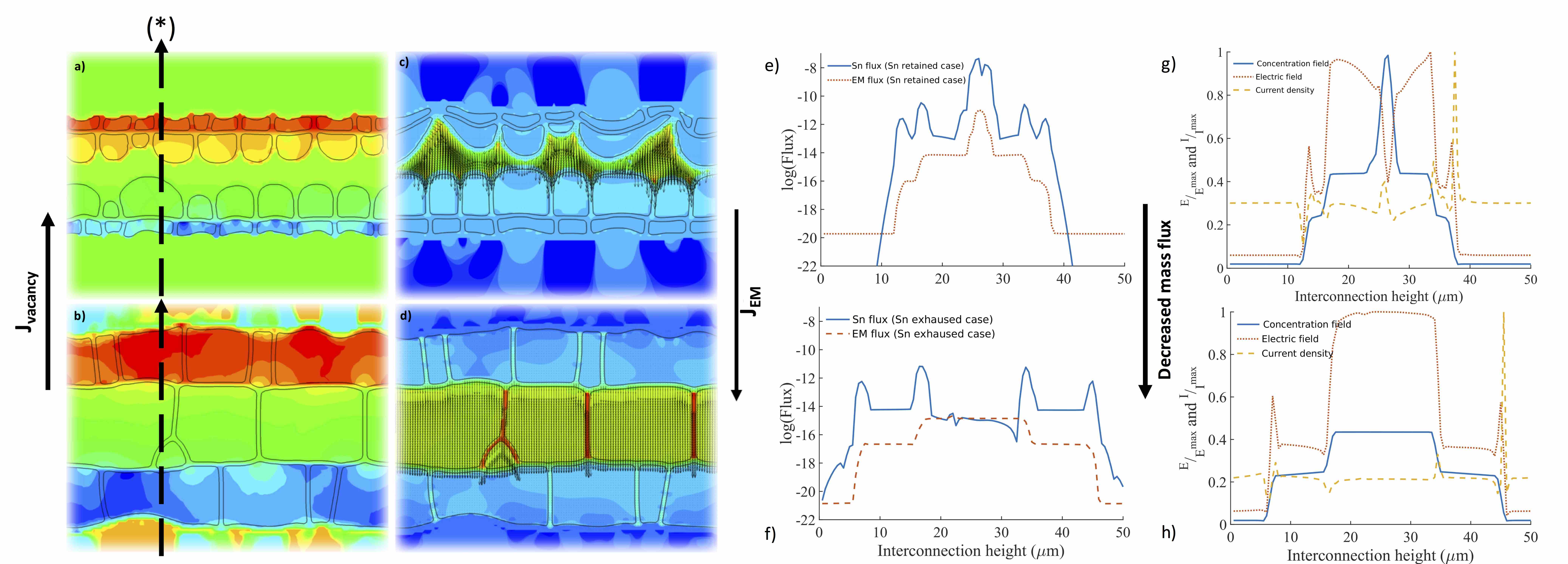}
\caption{ (a and b) Distribution of non-equilibrium vacancies in the microstructures emphasizing the role of the $Cu_6Sn_5$/$Cu_3Sn$ interface on generation of the vacancies at the anode and annihilation at the cathode layer. (c and d) EM flux in the microstructures. (e and f) EM and total mass flux along line (*) for Sn-retained and Sn-exhausted microstructures. (g and h) Normalized current density and electric field with corresponding concentration along line (*). The EM conditions are 180$^\circ$C and 4$\times$$10^{7} \sfrac{A}{m^2}$.} 
\label{fig:uniform_vac}
\end{figure*}

The same microstructures that are used in section \ref{sec:elec_res} are restudied for the evolution of the non-equilibrium vacancies. Figures~\ref{fig:uniform_vac}(a) and~\ref{fig:uniform_vac}(b) illustrate the distribution of non-equilibrium vacancies in the microstructures under 4$\times 10^7$ $\sfrac{A}{m^2}$ at 180$^\circ$C. Assuming a uniform vacancy distribution at the beginning, the vacancy migration in the opposite direction of the electron wind toward the cathode layer is shown in this figure. The direction of this flux is indicated by the arrow on the left hand side of this figure. With the unidirectional current flow, vacancy accumulation occurs at the blocking boundaries and depletion occurs at the other areas by means of severe difference in the diffusive ranges in the components/phases of the microstructure. 

%%%%%%%%%%%%%%%
%%%%%%%%%%%%%%%
%%%%%%%%%%%%%%%

%%%%%%%%%%%%%%%%
%%%%%%%%%%%%%%%%
%%%%%%%%%%%%%%%%

On the other hand, figs.~\ref{fig:uniform_vac}(c) and~\ref{fig:uniform_vac}(d) depict the distribution of EM fluxes in the microstructures indicating a severe EM flux toward anode in the interlayer phases of both microstructures. The quivers are indicating the location, direction and extent of this high flux. Hence, a considerable EM-mediated flux in the reverse direction of flux of vacancies is present in the Sn phase of Sn-retained microstructure. On the other hand, there is a considerable EM flux in the GBs of the Sn-exhausted microstructure which are parallel to the current direction. The vacancy evolution trend agrees with the theoretical vacancy evolution in the literature where an increase in the vacancy amount in the $Cu_{3}Sn$ layer at the cathode side is expected (red region in figs.~\ref{fig:uniform_vac}(a) and \ref{fig:uniform_vac}(b)). On the other hand, the vacancy amount in the $Cu_{3}Sn$ layer at the anode side decreases (blue region in the same figures). Figures~\ref{fig:uniform_vac}(e) and \ref{fig:uniform_vac}(f) illustrate the amount of total mass flux along the line (*) for the respective microstructures shown in fig.~\ref{fig:uniform_vac}(a) and \ref{fig:uniform_vac}(b). The change on the order of EM-mediated mass flux from TSV to the Sn phase (interlayer) is more than fourteen orders of magnitude in the Sn-retained structure. However, the flux only changes about eight orders of magnitude in the Sn-exhausted microstructure. In the mean time, there is an increase in the amount of current density in the Cu/$Cu_3Sn$ interface of cathode layer with respect to other interfaces as shown in figs.~\ref{fig:uniform_vac}(g) and \ref{fig:uniform_vac}(h). In the case where the applied current density is 4$\times 10^7$ $\sfrac{A}{m^2}$, the maximum current densities are 1.3$\times 10^8$ $\sfrac{A}{m^2}$ and 1.9$\times 10^8$ $\sfrac{A}{m^2}$ for Sn-retained and Sn-exhausted microstructures, respectively.

The constituent elements of binary IMCs usually have large differences in both size and electronegativity. As in this case, the atomic radius of Sn and Cu are very different, while the electronegativity is close. In Pauling's list of electronegativity, the values for Cu and Sn are 1.9 and 1.96, respectively. When the electron transfer takes place, the size of the atoms will change towards an ideal ratio. In the case of IMC, the electron being transferred from the hyperelectronic element (more electronegative) to the hypoelectronic element, causing the atomic volume to shrink \cite{pauling1987influence}. This charge transfer affects the local electron density and moves ions in the direction of the transfer, leaving vacancy spots which should be soon filled by other atoms. Otherwise, the process can lead to large clusters of vacancies and ultimate formation of macroscopic voids. EM mediated diffusion also accelerates the vacancy jumps in reverse direction of the current and the rate of this jump is directly proportional to Atomic Packing Factor (APF) of the crystal structure of the respective phases. The APF of monoclinic $Cu_6Sn_5$ and orthorhombic $Cu_3Sn$ are 0.707 and 0.764, respectively. The high value of APF in $Cu_3Sn$ indicates a possible slowdown in the migration of vacancies inside this IMC with respect to $Cu_6Sn_5$.

The experimental observations confirm the formation of voids after certain amount of time in the $Cu_3Sn$ IMCs of the cathode layer. This whole scenario plus the calculations in the previous section suggest that the migration of Cu atoms from cathode side to the anode side will leave vacancies behind. However, it is important to consider the rate of such a migration. Obviously, the rate is not consistent in the entire path and changes from one phase to the other one. In addition, the interfaces play different roles in this path, as they act as generators or annihilators of vacancies. This makes such a transport mechanism very complex. 

While the experiments suggest that the short-circuit channels and their quantity play important role as vacancy sink locations in prevention of the formation of Kirkendall voids, the distinct role of each of these channels are rarely addressed in the literature. This study illustrates that the horizontal $\sfrac{Cu_{6}Sn_{5}}{Cu_3Sn}$ interface act as the vacancy generation source in the cathode layer, while the same interfaces in the anode side acts as a vacancy annihilation source. On the other hand, other vertical interfaces and the GBs serve as the rapid pathways for the migration of atoms along the direction of the applied current. 

%%%%%%%%%%%%%%%%%%%%%%%%%%%%%%%%%%%%%%%%%%%%%%%%%%%%%%%
%%%%%%%%%%%%%%%%%%%%%%%%%%%%%%%%%%%%%%%%%%%%%%%%%%%%%%%
%%%%%%%%%%%%%%%%%%%%%%%%%%%%%%%%%%%%%%%%%%%%%%%%%%%%%%%

\section{Conclusions}

A multi-phase-field model for the joint formation and microstructure evolution in Cu/Sn/Cu joint of 3DIC systems is coupled with the equation of the continuity of charge and vacancy transportation to simulate major physical phenomena during the evolution of the interconnections. The particular characteristics of the morphological evolution are determined to be the fast transient in the establishment of the final equilibrium and are largely governed by the short-circuit diffusive channels, the characteristics of the interlayer phase, and the overall jump rate of the vacancies. In the presence of liquid Sn, the growth of the IMC layers is proportional to $d^n$ with $0.2 < n < 0.3$, and after exhaustion of the entire Sn (liquid) phase by the IMCs, the Cu-rich IMCs grow at the expense of each other by a slight deviation from the parabolic growth law. This growth rate is neither affected by the amount of solder material nor by the area fraction of the Cu substrate, and it is mostly dominated by the diffusive characteristics such as amount of fast diffusive channels and interfaces.

The sub-micron sized interconnection joints in the 3DIC systems are very narrow lines (thickness $<$ 15 $\mu m$) and are mostly composed of the IMCs due to successive reflow conditions. The horizontal interfaces along each IMC layer of these interconnections have a great role in creating diverging flux rates and ultimately shaping the growth pattern of the system. Under EM conditions, the low angle GBs of $Cu_{6}Sn_{5}$ IMCs get oriented toward the current direction and the interconnection forms a bamboo-shaped microstructure. Also, the locations where one type of IMC layer meets the other type are sites of a large divergence in the atomic flux. Atoms deplete at the upwind end of a polygranular cluster and accumulate at the downwind end, giving rise to tensile and compressive stresses, respectively.

Our results show that the interconnection covered with IMCs with grain sizes comparable to its structural dimension (e.g., widths) is less prone to the EM-induced changes in the structure. The dissolution of Cu atoms and hence the shrinkage in the cathode IMC layer or the faster growth in the anode IMC layer is the direct consequence of the rival fluxes in these interconnections under severe EM conditions.~Also, the concentration of non-equilibrium vacancy complexes in $Cu_3Sn$ increases overtime under EM conditions due to different diffusive characteristics of the Cu/$Cu_3Sn$/$Cu_6Sn_5$ layers.~A solution to this problem can be investigated by forming a one grain IMC phase of certain orientation and mechanical properties in between the Cu TSVs.~While the integrated computational framework developed in this study addresses many of the challenging circumstances in the 3DIC interconnections, the calculation of residual stresses upon evolution of the microstructure is also necessary.~Overall, to overcome the diffusion-driven issues in Cu/Sn/Cu interconnections, we suggest that the amount of interfaces that are perpendicular to the current direction should be significantly reduced to allow easier flux of Cu and Sn atoms in the joint structure. In addition, constructing highly coherent interfaces by growing secondary phases over the TSVs in the desired orientations may also result in less diffusion-driven failures due to enhanced flux capabilities.\\

\textbf{Acknowledgments}

The authors would like to acknowledge the ADA supercomputing facility of Texas A\&M University for providing computing resources useful in conducting the research reported in this paper. This research was supported by the National Science Foundation under NSF Grant No. CMMI-1462255.\\

\textbf{References}
\bibliography{Paper_acta}

\end{document}